\documentclass[preprint]{JASA}
\usepackage{float} 

\begin{document}
	\nolinenumbers
	~\\
	\begin{center}
		\textbf{\textcolor{blue}{ This manuscript has been submitted to the Journal of the Acoustical Society of America (JASA) and is currently under minor revision. Please note that subsequent versions may differ from this preprint.}}
	\end{center}

	\title[JASA/preprint]{Eardrum sound pressure prediction from ear canal reflectance based on the inverse solution of Webster's horn equation}
	\author{Reinhild Roden}
	\email{reinhild.roden@jade-hs.de}
	\author{Tobias Sankowsky-Rothe}
	\affiliation{Institut für Hörtechnik und Audiologie, Jade-Hochschule, Oldenburg, Germany.}
	
	\author{Nick Wulbusch}
	\altaffiliation{Also at: Institut für Mathematik, Carl von Ossietzky Universität Oldenburg, Germany.}
	\affiliation{Department für Medizinische Physik und Akustik, Carl von Ossietzky Universität Oldenburg, Germany.}
	
	\author{Alexey Chernov}
	\affiliation{Institut für Mathematik, Carl von Ossietzky Universität Oldenburg, Germany.}
	
	\author{Matthias Blau}			
	\altaffiliation{Also at: Cluster of Excellence ``Hearing4All''.}
	\affiliation{Institut für Hörtechnik und Audiologie, Jade-Hochschule, Oldenburg, Germany.}

	\preprint{Reinhild Roden, JASA}	
	
	\date{\today}

	\begin{abstract}
		\nolinenumbers
To derive ear canal transfer functions for individualized equalization algorithms of in-ear hearing systems, individual ear canal models are needed. In a one-dimensional approach, this requires the estimation of the individual area function of the ear canal. The area function can be effectively and reproducibly calculated as the inverse solution of Webster’s horn equation by finite difference approximation of the time domain reflectance. Building upon previous research, the present study further investigates the termination of the approximation at an optimal spatial resolution, addressing the absence of higher frequencies in typical ear canal measurements and enhancing the accuracy of the inverse solution. Compared to the geometric reference, more precise area functions were achieved by extrapolating simulated input impedances of ear canal geometries up to a frequency of 3.5\,MHz, corresponding to 0.1\,mm spatial resolution. The low pass of the previous work was adopted but adjusted for its cut-off frequency depending on the highest frequency of the band-limited input impedance. Robust criteria for terminating the area function at the approximated ear canal length were found. Finally, three-dimensional simulated and measured ear canal transfer impedances were replicated well employing the previously introduced and herein validated one-dimensional electro-acoustic model fed by the area functions.
		
	\end{abstract}
	

	\maketitle
	

	\section{\label{sec:intro} Introduction}
	\nolinenumbers
	Acoustic transparency is often desired to improve the sound quality of in-ear hearing systems. Transparency means that the device adjusts the sound pressure at the eardrum so that the user experiences a natural, open sound while wearing the hearing device. 
	To achieve this, the sound pressure at the eardrum with and without the hearing device must be known, ideally for each individual.
	The first step is to determine the transfer function of the individual ear canal when the device is in use, which is the focus of the present study.
	To this end, a hearing system equipped with a microphone placed at the residual ear canal can be employed as an impedance probe to derive the input impedance of the residual ear canal from a short measurement \cite{sankowsky2011prediction} \cite{vogl2019individualized}. The ear canal input impedance provides sufficient information to potentially assess the sound pressure at the eardrum. 
	There are various methods to predict sound pressure at the eardrum, but many of them \cite{lewis2015non} \cite{vogl2019individualized} \cite{wulbusch2023using} involve complex fitting processes and associated challenges, including high computational demands, the need for carefully selecting fitting parameters (e.g., the number of iterations, error tolerances, initial values, constraints) or convergence problems due to local minima of the error function. 
	Some may also require additional knowledge, such as the approximate ear canal length in order to constrain the parameter space.\\
	Hence, a single concise calculation like the inverse solution of Webster's horn equation using a finite-difference approximation from time domain reflectance, as proposed by \citet{rasetshwane2011inverse} with a subsequent one-dimensional transfer model would be preferable. However, as will be shown below, the original method in \citet{rasetshwane2011inverse} unfortunately underestimates area functions, which means that the predicted sound pressure at the eardrum may be inaccurate.\\
	Considering the analytical case of an exponential horn, a parabolic horn and a conical horn with known monotonically growing shapes, the method in \citet{rasetshwane2012reflectance} differs from \citet{rasetshwane2011inverse} in that the spatial resolution is much higher and the number of iterations limiting the finite difference approximation is given by the known length of the horn. Inverse solutions were shown to agree very well with the theoretical ones, but real ear canals may pose additional challenges. 
	Related literature mainly refers to the problem of a non-casual time domain reflectance due to neglecting wavefront curvature and evanescent modes for non-uniform ear canals \cite{norgaard2019calculation} \cite{keefe2020sound}.
	Apart from these concerns, a consistent investigation concerning the dependence of prediction accuracy on selected parameters of the method would be a useful supplement. Furthermore, to the best of our knowledge, there is currently no validation comparing the derived area functions of individual ear canals with a known reference from three-dimensional scans, nor regarding resulting eardrum pressure predictions.\\
	The present study aims to modify the method by \citet{rasetshwane2011inverse} while retaining the model assumptions to further enhance the prediction accuracy of the area function of individual ear canals and the eardrum sound pressure. The paper is organized as follows: 
	Section~\ref{sec:drum_pressure_calc} provides a comprehensive overview of the theoretical foundations that have been established in previous studies, needed to calculate the drum pressure using the inverse solution of Webster's horn equation by a finite difference approximation.
	The methods employed in the present study are delineated in Section~\ref{sec:methods}, which is further subdivided into two parts.
	Firstly, the reference data on acoustic transmission in ear canals are introduced. The second part contains the procedure and methodological modifications to previous work for the best possible estimation of the sound pressure at the eardrum.
	In section~\ref{sec:results}, the dependence of the prediction accuracy on selected parameters is investigated by comparing the estimate of the ear canal transfer impedance to three-dimensional finite element method (FEM) simulations. After selecting the best parameters, the optimized method is validated on the basis of measurements. This is followed by the discussion and conclusion in Sections~\ref{sec:discussion} and \ref{sec:conclusion}.
	
	\newpage
	\section{\label{sec:drum_pressure_calc} Drum pressure calculation via inverse solution of Webster's horn equation}
	
	\subsection{\label{sec:reflectance} Reflectance at the entrance of horns}
	The input impedance at the lateral end of an ear canal, $Z\rm{_{ec}(\omega)}$, with the angular frequency~$\omega$, can be determined by measuring the sound pressure at the lateral end with known source parameters and microphone characteristics \cite{keefe1992method} \cite{voss1994measurement} \cite{blau2010prediction} \cite{sankowsky2011prediction} \cite{sankowsky2012prediction} \cite{sankowsky2015individual} \cite{vogl2019individualized} \cite{rasetshwane2011inverse}.
	The frequency domain reflectance $R(x,\omega)$ at the lateral end $x=0$, with $x$ giving the center axis position, results from the input impedance $Z\rm{_{ec}}$ and the characteristic impedance $Z\rm{_0}$ at $x=0$,
	\begin{eqnarray}
		{R}\left(x=0,\omega \right)=\frac{{Z}_{ec}\left(\omega \right)-{Z}_{0}\left(x=0\right)}{{Z}_{ec}\left(\omega \right)+{Z}_{0}\left(x=0\right)}~.
		\label{eq:R}
	\end{eqnarray}
	In general, $Z\rm{_{0}}$ is the ratio of the fluid density $\rho$ times the speed of sound $c$ and the area function $A$,
	\begin{eqnarray}
		Z\rm{_{0}}\left(\textit{x}=0\right)=\frac{\rho \textit{c}}{A\left(\textit{x}=0\right)}\,,
		\label{eq:Z0}
	\end{eqnarray}
	which implies that the area function at x=0 must be known. Here, we follow
	the iterative procedure for estimating the characteristic impedance $Z\rm{_0}(x=0)$ in \citet{rasetshwane2011inverse}, by using MATLAB scripts provided in \citet{rasetshwaneSourceCode2012old} for measurements on unknown ear canals. The steps of this procedure are outlined in Section~\ref{sec:procedure_of_previous_work}.
	It should be added that Eq.~(\ref{eq:Z0}) only applies to ideal acoustic systems with a purley real-valued characteristic impedance $Z\rm{_0}$ without dispersion or any damping. In real systems like ear canals, it may also contain an uncertain and presumably negligible imaginary part describing losses and phase shifts (see e.g. \cite{kuttruff2007acoustics}, p. 155, regarding dispersion in exponential horns).
	
	\subsection{\label{sec:inversesolution} Inverse solution of Webster's horn equation}
	Webster's horn equation was derived for semi-infinite and rigidly bounded horns with slowly varying area functions, for frequencies with large wavelengths compared to the transversal horn dimension to ensure one-dimensional propagation \cite{webster1919acoustical}. Applications often address horns of monotonically growing area functions as in the phonograph or acoustic loudspeakers \cite{goldsmith1924performance}. The equation can also be applied to horn structures in microphones to optimize acoustic performance and enhance frequency response, to horn channels in hearing aids, to horn shaped structures in auditory spaces or concert halls, in musical instruments such as trumpets or in acoustic sensors.\\
	Following the notation in \citet{rasetshwane2011inverse} and \citet{rasetshwane2012reflectance}, Webster's horn equation is given by
	\begin{eqnarray}
		{\partial }_{x}p(x,t)=-\frac{\rho }{A\left(x\right)}{\partial }_{t}u(x,t)
		\label{eq:WH1}
	\end{eqnarray}
	and 
	\begin{eqnarray}
		{\partial }_{x}u(x,t)=-\frac{A\left(x\right)}{\rho {c}^{2}}{\partial }_{t}p(x,t)~,		\label{eq:WH2}
	\end{eqnarray}
	where $p(x,t)$ is the sum of the superimposed incoming $p_{+}\left(x,t\right)$ and outgoing wave $p_{-}\left(x,t\right)$,
	\begin{eqnarray}
		p\left(x,t\right)=p_{+}\left(x,t\right)+p_{-}\left(x,t\right)\,.
	\end{eqnarray}
	The time domain reflectance $r\left(x,t\right)$ is defined as the deconvolution of $p_{-}(x,t)$ by $p_{+}(x,t)$. It is assumed to be zero for $t<0$ for all values of $x$, ensuring causality. 
	Eliminating the volume velocity $u(x,t)$ in Eqs.~(\ref{eq:WH1}) and (\ref{eq:WH2}) results in an equation of second order,
	\begin{eqnarray}
		{\partial }_{x}^{2}p(x,t)+2\epsilon \left(x\right){\partial }_{x}p(x,t)-\frac{1}{{c}^{2}}{\partial }_{t}^{2}p(x,t)=0\,,
	\end{eqnarray}
	with 
	\begin{eqnarray}
		\epsilon \left(x\right)\equiv \frac{1 }{2}\frac{d}{dx}\mathrm{ln}A\left(x\right)~.
	\end{eqnarray}
	The entire procedure for solving Webster's horn equation for $A(x)$ by finite difference approximation is described by \citet{rasetshwane2012reflectance} and a MATLAB implementation was provided in \citet{rasetshwaneSourceCode2012old} and \citet{rasetshwaneSourceCode2012}.\\
	By means of the inverse Fourier transform of $R(x,\omega)$, the time domain reflectance $r(x,t)$ is obtained. In discrete time, the temporal resolution $\Delta t$ of $r(x,t)$ must be chosen small enough to satisfy the stability criterion according to \citet{courant1928partiellen}. Here, this is ensured by linking $\Delta t$ to the highest frequency $f\rm{_{sup}}$ with $\Delta t=1/f\rm{_{sup}}$.
	\subsection{\label{sec:procedure_of_previous_work} Previous work}
\begin{figure}[H]
		\centering

		\figline{
			
			\includegraphics[trim={0cm 0.5cm 0cm 0cm},width=0.5\linewidth]{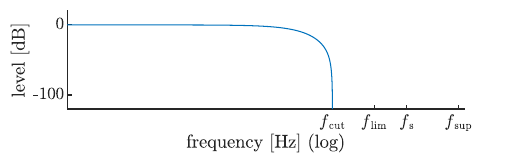}	
			
		}
		
		\caption{Blackman window as designed by \citet{rasetshwane2011inverse} and also found in \citet{rasetshwaneSourceCode2012}, and characteristic frequencies.}
		\label{fig:BW}
	\end{figure}

	For the present study, there are two previous publications regarding the inverse solution. The first article \cite{rasetshwane2011inverse} is dedicated to the inverse solution of Webster's horn equation using a finite difference approximation to time domain reflectance, for measurements with a fixed sampling rate $f\rm{_s}$\,=\,48\,kHz on human ear canals. The authors used upsampling at rates up to 192 kHz, by zero-padding the frequency domain reflectance. The upsampling rates $f\rm{_{sup}}$ were obtained from integer multiples $n\rm{_{sup}}$ of the sampling rate $f\rm{_s}$. They also used an amplitude correction by multiplying the frequency domain reflectance with $n\rm{_{sup}}$ as found in the earlier version of the supplementary material \cite{rasetshwaneSourceCode2012old}. This correction is no longer found in the later version \cite{rasetshwaneSourceCode2012}. A frequency domain Blackman window $W(\aleph)$ given by
	\begin{eqnarray}
		W(\aleph) = \left\{
		\begin{array}{ll}
			\frac{1-a+\cos(\aleph)+a\cdot \cos(2\cdot \aleph)}{2}  & \mbox{if $0\le\aleph\le\pi$} \\
			0 & \mbox{if $\aleph>\pi$}
		\end{array}
		\right.
		\label{eq:BW0}
	\end{eqnarray}
	for
	\begin{eqnarray}	
		\aleph=\frac{\pi \cdot n\cdot f_{sup}}{N\cdot f_{cut}}\,,
		\label{eq:BW3}
	\end{eqnarray}
	$a$\,=\,0.16 and $n$\,=\,0,1,...,$N$/2, where $N$ is the FFT length, was used on the upsampled frequency domain reflectance (conforming with \citet{rasetshwaneSourceCode2012old} and \citet{rasetshwaneSourceCode2012}). This window increasingly attenuated levels at higher frequencies and suppressed them completely from $f\rm{_{cut}}$ onwards, by setting $W(\aleph>\pi)=0$. Fig.~\ref{fig:BW} shows this frequency domain window, together with all other specific frequencies of the algorithm: $f\rm{_{lim}}$ defines the highest frequency for which a value for the input impedance or reflectance is available. In \citet{rasetshwane2011inverse}, $f\rm{_{lim}}$ was set to the Nyquist frequency $f\rm{_s}$/2 and  $f\rm{_{cut}}$ was smaller than $f\rm{_{lim}}$.\\
	Lastly, the TDR for $t$\,$>$\,0 was reversed in time and added with the reflectance for $t$\,$>$\,0 (time-reversed addition), eliminating a signal-processing artifact not reported in detail. This step has disappeared in \citet{rasetshwaneSourceCode2012}.

	As previously mentioned in Section \ref{sec:reflectance}, \citet{rasetshwane2011inverse} adjusted $Z_0$ by an iterative procedure reducing the TDR at $t=0$, which unfortunately was not explained in more detail. The supplementary material \cite{rasetshwaneSourceCode2012old} contains the source code for the aforementioned process, henceforth referred to as 'surge~I'. In the first step, an initial estimate of $Z\rm{_0}(\textit{x}=0)$ is calculated. In the second step, the frequency domain reflectance $R(x=0,\omega)$ is calculated, using the current estimate of $Z\rm{_0}$, where the frequency domain Blackman window is used to eliminate ringing in the TDR.
	Thirdly, the mean real part of the current windowed $R(x=0,\omega)$, denoted by $m_{R,k}$, is divided by the mean of the window function $m_W$, and $Z_{0,k}$ is iteratively updated via 
	\begin{eqnarray}
		Z_{0,k+1} = Z_{0,k} \cdot (1+\frac{m_{R,k}}{m_W})\,.
		\label{eq:surge}
	\end{eqnarray}
	Steps two and three are repeated until the value of $Z\rm{_0}(x=0)$ converges to a single value.\\
	A modified variant adjusting the characteristic impedance can be found in \citet{rasetshwaneSourceCode2012}, henceforth referred to as 'surge II'. In this approach, $R(x=0,\omega)$ was not calculated with the real-valued $Z_0(\textit{x}=0)$ as in Eq.~(\ref{eq:R}), but rather with a complex-valued characteristic impedance.\\
	Subsequent to the calculation of $Z\rm{_0}(x=0)$ and $R(x=0,\omega)$, the inverse solution can be calculated.
	Since the correct length of the ear canal is not known initially, considerations for abort criteria or constraints for the finite difference approximation are needed. Firstly, \citet{rasetshwane2011inverse} set a maximum length $l\rm{_{max}}$, which was assumed to be larger than the expected geometric length $l\rm{_{geoend}}$ of the unknown ear canal or simulator. Subsequently, the TDR peak latency was associated with the length $l\rm{_{TDRmax}}$ at the point of reflection at the eardrum. Furthermore, it was assumed in \citet{rasetshwaneSourceCode2012} that the possible end of the entire ear canal is at length $l\rm{_{TDR50}}$ where the TDR has decayed to 50\% after reaching its maximum.\\
	\citet{rasetshwane2011inverse} mentioned to validate their method by the use of an ear canal simulator consisting of 15 brass tubes connected to each other, with a rigid termination. The best match occurred for an upsampling factor $n\rm{_{s}}$ of 4 at a sampling rate of 48\,kHz with $f\rm{_{cut}}$\,=\,17\,kHz, see Fig.~2 in \citet{rasetshwane2011inverse}.\\
	In addition, the spatial resolution $\Delta x$ is quite coarse with $\approx$\,1.8\,mm for $f\rm{_{sup}}$\,=\,4$\cdot$48\,kHz. In contrast, a spatial resolution of the ear canal of at least 1\,mm is recommended by \citet{hudde1999methods} and even 0.1\,mm by \citet{xia2024effect}. 
	Thus, there are four parameters in the work of \citet{rasetshwane2011inverse}, namely $f\rm{_{lim}}$=$f\rm{_s}$/2, $f\rm{_{cut}}$, $f\rm{_{sup}}$ and the length $l$ terminating the ear canal, that need to be considered in more detail. A consistent exploration of the dependency of the resulting $Z\rm{_{trans}}$ on selected parameters is therefore undertaken in the present work.\\
	For geometries as analyzed in the second article \citep{rasetshwane2012reflectance}, the impedance or reflectance can also be analytically determined. Authors considered simply shaped exponential, conical and parabolic horns with known monotonically growing area functions and a priori known lengths. In \citet{rasetshwane2012reflectance}, the method from \citet{rasetshwane2011inverse} was used but the chosen sampling rate of 1\,MHz with accompanying spatial resolution $\Delta$\,$x$\,$\approx$\,0.35\,mm made upsampling and low-pass filtering unnecessary. Eq.~(\ref{eq:Z0}) with the known $A(x=0)$ was employed to calculate $Z\rm{_0}(x=0)$ without any adjustment. The inverse solution resulted in quite accurate area functions compared to the known reference with diameter deviations less than 4\,$\%$. As discussed by the authors, remaining errors might be caused by the violated assumption of one-dimensional wave propagation for smaller wave lengths and the incorrect assumption for the characteristic impedance especially for parabolic horns. Furthermore, authors reported that errors systematically increase for smaller sampling rates, which also addresses the issue of band-limited measurements on ear canals.\\

	\subsection{\label{sec:EAmod} One-dimensional electro-acoustic model}
\begin{figure}[H]
		\baselineskip=12pt
		\centering
		\includegraphics[width=.4\textwidth]{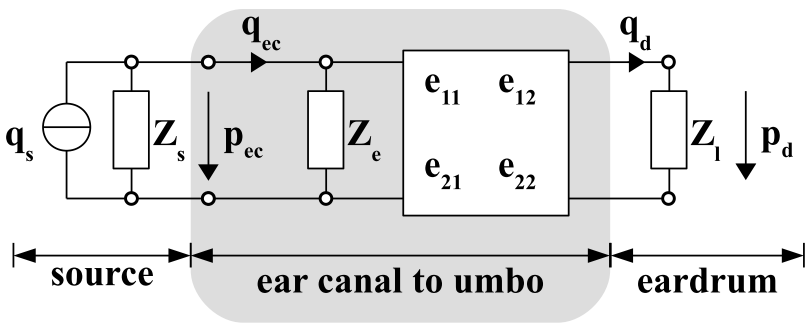}
	\caption{One-dimensional electro-acoustic model used to calculate the transfer impedance $Z\rm{_{trans}}$ from the input impedance $Z\rm{_{ec}}$ of the ear canal (defined as ratios of the drum pressure $p\rm{_{d}}$ or the pressure at the lateral end of the ear canal $p\rm{_{ec}}$ to the volume velocity at the lateral end of the ear canal $q\rm{_{ec}}$, Eqs.~(\ref{eq:Zec}) and (\ref{eq:Ztrans_pd}).}
		\label{fig:EA_model}
	\end{figure}

The estimated area function given by the inverse solution of Webster's horn equation is only an intermediate step. In order to compute the eardrum sound pressure $p\rm{_d}$, a one-dimensional approach via two-port models comprising a source model, a transmission model and a load model as introduced by \citet{sankowsky2015individual} is used in the present study. The circuit diagram of this electro-acoustic (EA) model is shown in Fig.~\ref{fig:EA_model}. 
The transmission matrix of the ear canal, whose parameters $e_{ij}$ are determined by the area function, results from a chain of single tube segments, each modeled as a lossless acoustic duct according to \citet{benade1968propagation}. 
The load $Z\rm{_l}$ represents the eardrum acoustic impedance acting in parallel to the acoustic impedance of the residual volume medial to the umbo point projected on the centerline, according to \citet{hudde1998measuringII}. The shunt impedance $Z\rm{_e}$ represents the residual error between the measured and modeled $Z\rm{_{ec}}$ and compensates for additional leakage, unmodeled attenuation and the individual mismatch between the true and generalized non-individual eardrum impedance. 
	
	Since the (known) input impedance $Z\rm{_{ec}}$ already contains information on the attenuation by thermo-viscous effects, skin and leakage, one can derive an expression for the transfer impedance $Z\rm{_{trans}}$ without explicit reference to $Z\rm{_l}$ and $Z\rm{_e}$. Using the circuit diagram from Fig.~\ref{fig:EA_model}, one obtains
	\begin{eqnarray}
		\left[\begin{array}{cc}p\rm{_{ec}}\\ q\rm{_{ec}}\end{array}\right]=\left[\begin{array}{cc}1&0\\ \frac{1}{{Z}\rm{_{e}}}&1\end{array}\right]\left[\begin{array}{cc}{e}_{11}&{e}_{12}\\ {e}\rm{_{21}}&{e}\rm{_{22}}\end{array}\right]\left[\begin{array}{c}{p}\rm{_{d}}\\ q\rm{_{d}}\end{array}\right]= \textbf{B} \left[\begin{array}{c}{p}\rm{_{d}}\\ q\rm{_{d}}\end{array}\right]\,.
		\label{eq:twoport}
	\end{eqnarray}

as the relationship between sound pressure and volume velocity at the lateral end of the ear canal, $p\rm{_{ec}}$ and $q\rm{_{ec}}$,  and the sound pressure and volume velocity at the eardrum, $p\rm{_{d}}$ and $q\rm{_{d}}$.
Noting that

\begin{eqnarray}	
		\textit{Z}\rm{_{ec}}=\frac{\textit{p}\rm{_{ec}}}{\textit{q}\rm{_{ec}}}\,,
		 \label{eq:Zec}
	 \end{eqnarray}
	\begin{eqnarray}	
		\textit{Z}\rm{_{trans}} = \frac{\textit{p}\rm{_d}}{\textit{q}\rm{_{ec}}}\,,
		\label{eq:Ztrans_pd}
	\end{eqnarray}
	it follows through the inverse matrix $\textbf{B}^{-1}$, whereas det(\textbf{B})\,=\,1 according to the reciprocity principle for passive linear systems \cite{strutt1871some}, that
	\begin{eqnarray}	
		{Z}\rm{_{trans}} = {e}\rm{_{22}}{Z}\rm{_{ec}} - {e}\rm{_{12}} + \frac{{e}\rm{_{12}}Z\rm{_{ec}}}{{Z}\rm{_{e}}}\,.
		\label{eq:Ztrans}
	\end{eqnarray}
	\citet{sankowsky2015individual} suggested neglecting the third term in Eq.~(\ref{eq:Ztrans}), as it is considered to be small. More specifically, $Z\rm{_e}$ is a mainly mass-dominated impedance with low level at low frequencies and increasing level towards high frequencies. Depending on the size of the leakage, the levels of $Z\rm{_e}$ and $Z\rm{_{ec}}$ cross at a frequency which is typically below 1\,kHz for smaller leakage due to venting or a slightly leaky fit of the ear mold. Hence, the third term becomes very small above 1\,kHz. At lower frequencies, it can be safely assumed that $e\rm{_{12}}$ is very low since in acoustically short waveguides it is approximately proportional to the ratio of the ear canal length to the wavelength.
	Thus, as given in \citet{sankowsky2015individual}, the transfer impedance is merely an expression of only two transfer parameters of the ear canal and the measured input impedance,
	\begin{eqnarray}	
		{Z}\rm{_{trans}} \approx {e}\rm{_{22}}{Z}\rm{_{ec}} - {e}\rm{_{12}}\,.
		\label{eq:Ztrans_approx}
	\end{eqnarray}
	Finally, if $q\rm{_{ec}}$ is computed using the in-ear-hearing device as probe to determine $Z\rm{_{ec}}$ \cite{blau2010prediction} \cite{sankowsky2015individual} \cite{vogl2019individualized}, the sound pressure at the eardrum $p\rm{_d}$ can be derived from $Z\rm{_{trans}}$ in Eq.~(\ref{eq:Ztrans_pd}). Errors in the prediction of $Z\rm{_{trans}}$ thus also result in errors on the predicted $p\rm{_d}$. In the following sections, the transfer impedance $Z\rm{_{trans}}$ is considered.\\
	\newpage
	\section{\label{sec:methods} Methods}
	\subsection{\label{sec:refdata} Reference data on acoustic transmission in ear canals}
	\subsubsection{\label{sec:refA} Geometrically determined area functions of ear canals}
\begin{figure}[H]
		\centering
		\baselineskip=12pt
		\figline{
			\includegraphics[trim={0cm 1.3cm 0cm 1.6cm},clip,width=.2\linewidth]{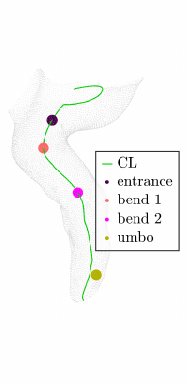}
			\includegraphics[trim={0cm 0cm 0cm 0cm},clip,width=.15\linewidth]{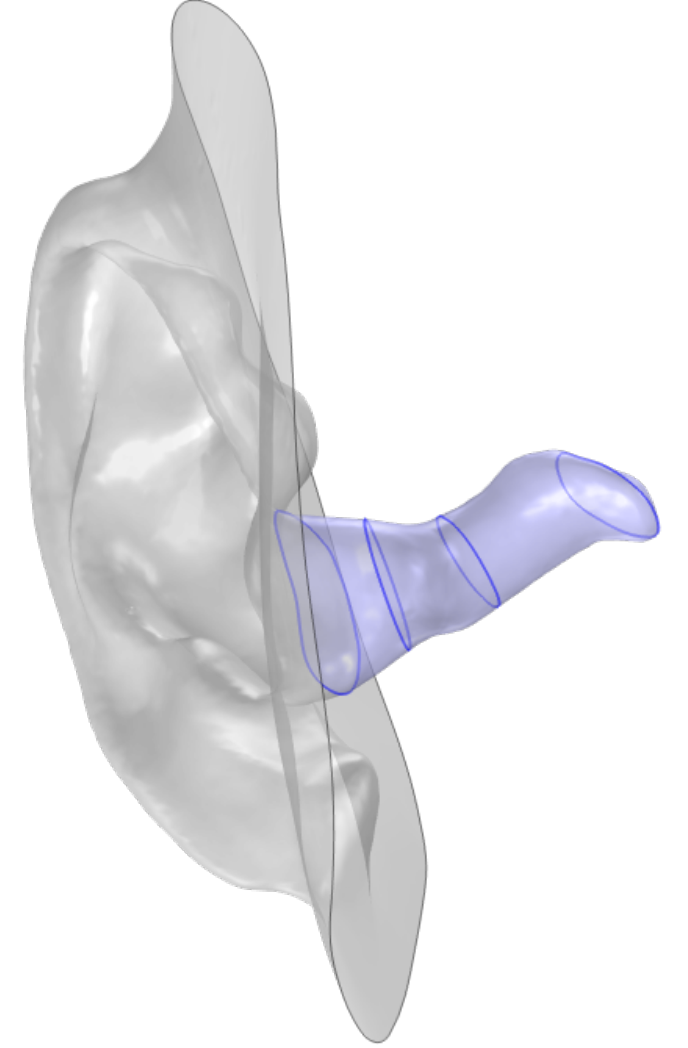}
		}
		
		\caption{Right outer ear of subject 3 in the IHA database with the extracted centerline and landmarks (left) and with final cut planes at the entrance, the first and the second bend and at the eardrum (right).}
		\label{fig:CL_example}
	\end{figure}

\begin{figure}[H]
		\centering
		\includegraphics[trim={0cm 0cm 0cm 0.2cm},width=.5\textwidth]{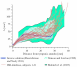}
		
		\caption{Area functions of right ear canals cut at the entrance plane for IHA database subjects 1 - 21, and corresponding data from \citet{rasetshwane2011inverse}, \citet{stinson1989specification}, \citet{balouch2023measurements}.}
		\label{fig:area_fun}
	\end{figure}

	The IHA database of individual three-dimensional geometries consisting of torso, head and complete outer ears including the ear canals with eardrums \cite{roden2020iha} (partly published in \citet{roden2021iha}) is being used in the present study. More specifically, the geometries of the right ear canal of subjects 1 to 21 in the IHA database were used. First, the ear canals (and parts of the concha) were extracted from the stl files in the database. The centerline (CL) of each ear canal between a point lateral to the ear canal entrance and the innermost corner of the ear canal was then determined utilizing the vmtk toolbox \cite{antiga2002patient}. CLs were also determined according to \citet{stinson1989specification}. Both methods resulted in similar CLs, except for strongly curved ear canals where the vmtk method appeared to better follow the shape of the ear canals. (A detailed comparison is beyond the scope of the present article.) In the following, CLs extracted with the vmtk toolbox are used for the geometric analysis.\\
	The entrance of the ear canal was defined as the first plane (seen from the lateral side) orthogonal to the CL  which intersected the ear canal walls to form a closed ring of polygon vertices, as proposed by \citet{balouch2023measurements}. The first and second bends were initially determined by visual inspection. The nearest local torsion or curvature maximum of the CL was then used to determine the first and second bend on the CL in a reproducible manner. For computations with ear canals of different lengths, the ear canals were sectioned by planes orthogonal to the CL at the entrance and at the first and the second bend. Fig.~\ref{fig:CL_example} shows an example ear canal geometry with CL and landmarks (left) and a pinna with resulting cut planes (right). The areas of the orthogonal cross-sections along the CL result in area functions shown in Fig.~\ref{fig:area_fun} for the right ear canals of subjects 1 to 21 of the IHA database (cut at the entrance), as well as for ear canals from various other studies.\\
	The surface of the tympanic membrane was identified by geometric estimation of its approximate rim. For this purpose, the cross-sectional polygons of the medial end of the ear canal were interpolated and the points with the maximum curvature in the region of the rim were extracted. The plane with the smallest Euklidian distance from these points was defined as cut plane to separate the ear canal walls and eardrum. The umbo point was identified as the highest elevation of the tympanic membrane above the cut plane.\\
	
	\subsubsection{\label{sec:sim} Three-dimensional FEM simulations of input and transfer impedances}
\begin{figure}[H]
		\centering
		\baselineskip=12pt
		\figline{\hfill\includegraphics[trim={0cm 0.2cm 0cm 0cm},clip,width=.3\textwidth]{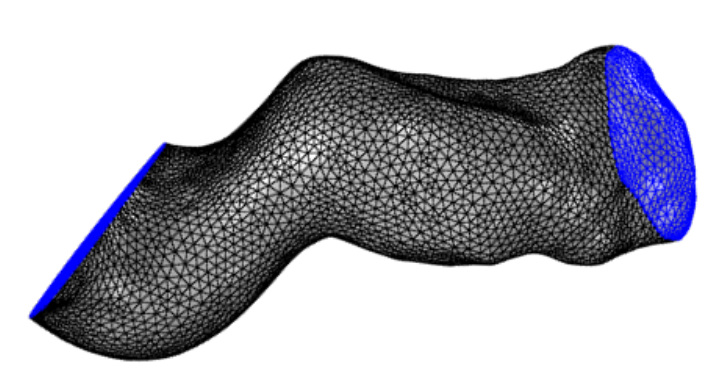}\label{fig:EC_mesh_entrance:a}
			\includegraphics[trim={0cm 0cm 0cm 0cm},clip,width=.25\textwidth]{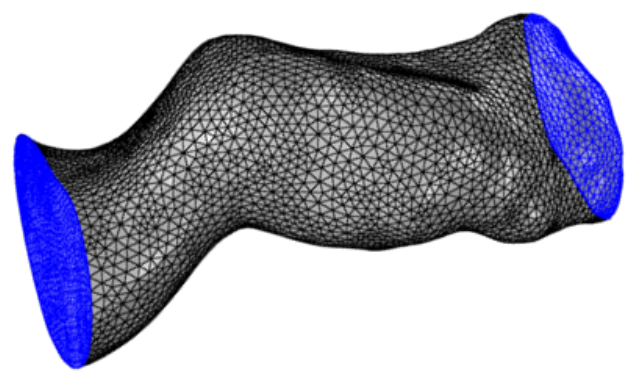}\label{fig:EC_mesh_bend1:b}
			\includegraphics[trim={0cm 0cm 0cm 0cm},clip,width=.16\textwidth]{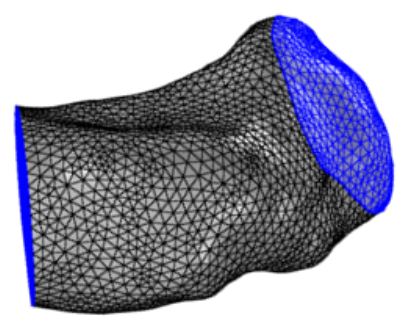}\label{fig:EC_mesh_bend2:c}\hfill}
		\caption{Mesh of the right residual ear canal for subject 5 of the IHA database cut at the entrance plane (left), at the first bend (mid), and at the second bend (right).}
		\label{fig:EC_meshes}
	\end{figure}

	\textbf{Simulations using human ear canal geometries}\\
	In order to analyze the input and transfer impedance of ear canals, three-dimensional FEM simulations were carried out utilizing COMSOL Multiphysics 6.2 (COMSOL AB, Stockholm, Sweden) in a similar way as described in \citet{roden2020iha} and \citet{wulbusch2023using}. The impedance according to \citet{hudde1998measuringII} for the tympanic membrane and medial structures was assigned to the eardrum surface.\\
	As suggested by \citet{berggren2018acoustic} and \citet{bach2018theory}, the thermo-viscous effects in the immediate vicinity of the ear canal walls were described by a boundary layer impedance (BLI) as a weak contribution to the Helmholtz equation. Furthermore, the skin impedance as provided by COMSOL, based on experimental data \cite{haakansson1986mechanical}, was assigned to the ear canal walls.\\
A normal velocity $v\rm{_{n,ec}}$ was applied to the lateral terminating surface $A\rm{_{ec}}$, which could be either at the entrance, the first or the second bend of the ear canal. The input impedance $Z\rm{_{ec}}$= $p\rm{_{ec}}$/$q\rm{_{ec}}$ was determined by the average sound pressure $p\rm{_{ec}}$ of all pressures $p$ on $A\rm{_{ec}}$ divided by the volume velocity $q\rm{_{ec}}$ on $A\rm{_{ec}}$,
\begin{equation}
	p\rm{_{ec}} = \frac{\int{}{}{\textit{p}} \mathrm{d}\textit{A}\rm{_{ec}}}
	{ \textit{A}\rm{_{ec}}	}\,,
	\label{eq:pec}
\end{equation}
and
\begin{equation}
	q\rm{_{ec}} = \int{  }{  }{  \textit{v}_{n,ec} } \mathrm{d}\textit{A}\rm{_{ec}}.
	\label{eq:qec}
\end{equation}
	The transfer impedance $Z\rm{_{trans}}$\,=\,$p\rm{_{d}}$/$q\rm{_{ec}}$ was based on $p\rm{_d}$ calculated at the umbo point. The simulation was performed for linearly distributed frequencies between 0.1 and 20\,kHz with a resolution of 100\,Hz.\\
	Sample meshes with excitation surfaces and tympanic membranes marked in blue are shown in Fig.~\ref{fig:EC_meshes}. In the left column of Fig.~\ref{fig:Zin_Ztrans}, resulting levels of simulated input and transfer impedances, as well as according level differences, are shown for all 21 ear canals, each cut at the entrance, at the first bend and at the second bend. The observed level increase towards low frequencies is characteristic for $Z\rm{_{ec}}$, as both the ear canal and the tympanic membrane act as compliances at low frequencies. The variation of the residual ear canal lengths and shapes are reflected by shifted levels as well as shifted minima and maxima in the magnitude of $Z\rm{_{ec}}$. The ratio of $Z\rm{_{trans}}$ to $Z\rm{_{ec}}$ is 0\,dB at low frequencies, as expected for large wavelengths. This is related to the fact that the transfer parameter $e\rm{_{12}}$ is small in level at low frequencies, see discussion in Section~\ref{sec:EAmod}. Anti-resonances of $Z\rm{_{ec}}$ are canceled in $Z\rm{_{trans}}$, whereas the frequencies of the resonances coincide.\\

\begin{figure}[H]
		\centering
		\baselineskip=12pt
		\figcolumn{\hfill	
			
			\includegraphics[trim={0.8cm 3cm 0.8cm 2.1cm},clip,width=.5\linewidth]{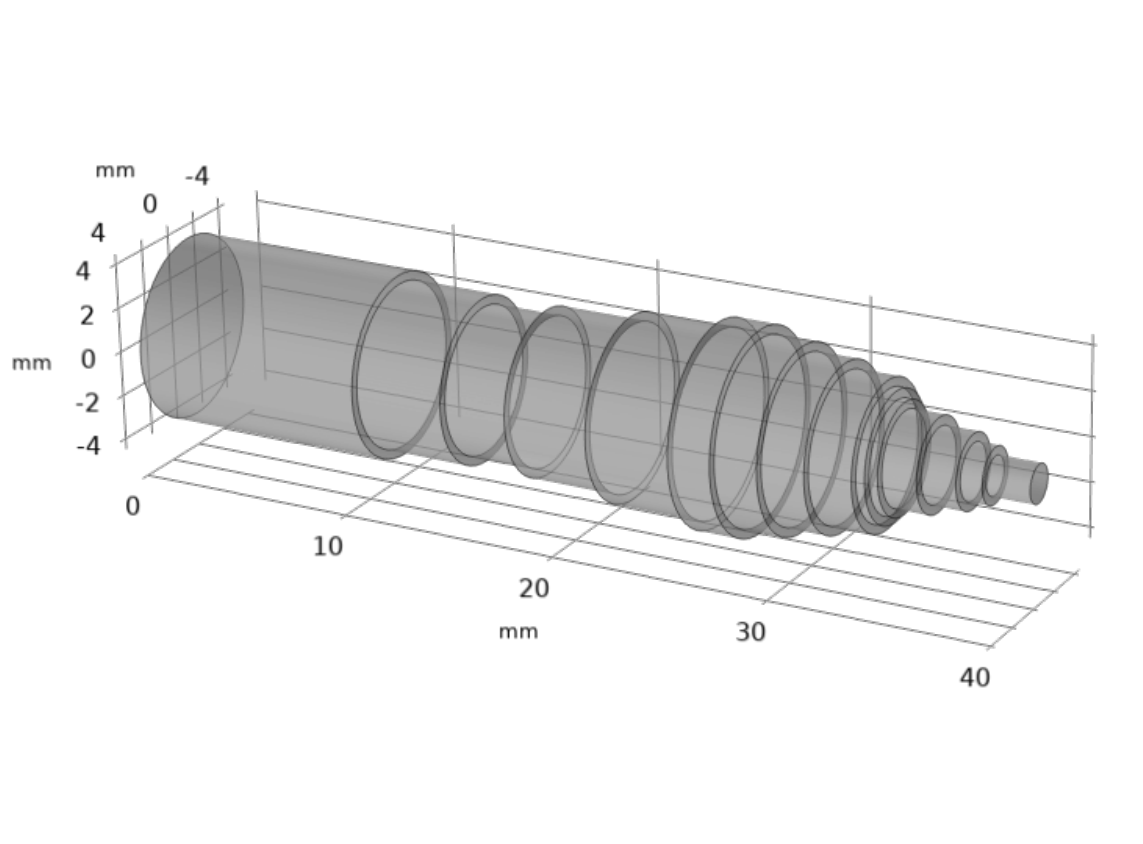}
			\figline{\hfill	
				\includegraphics[trim={1cm 3.5cm 0cm 2.1cm},clip,width=.5\linewidth]{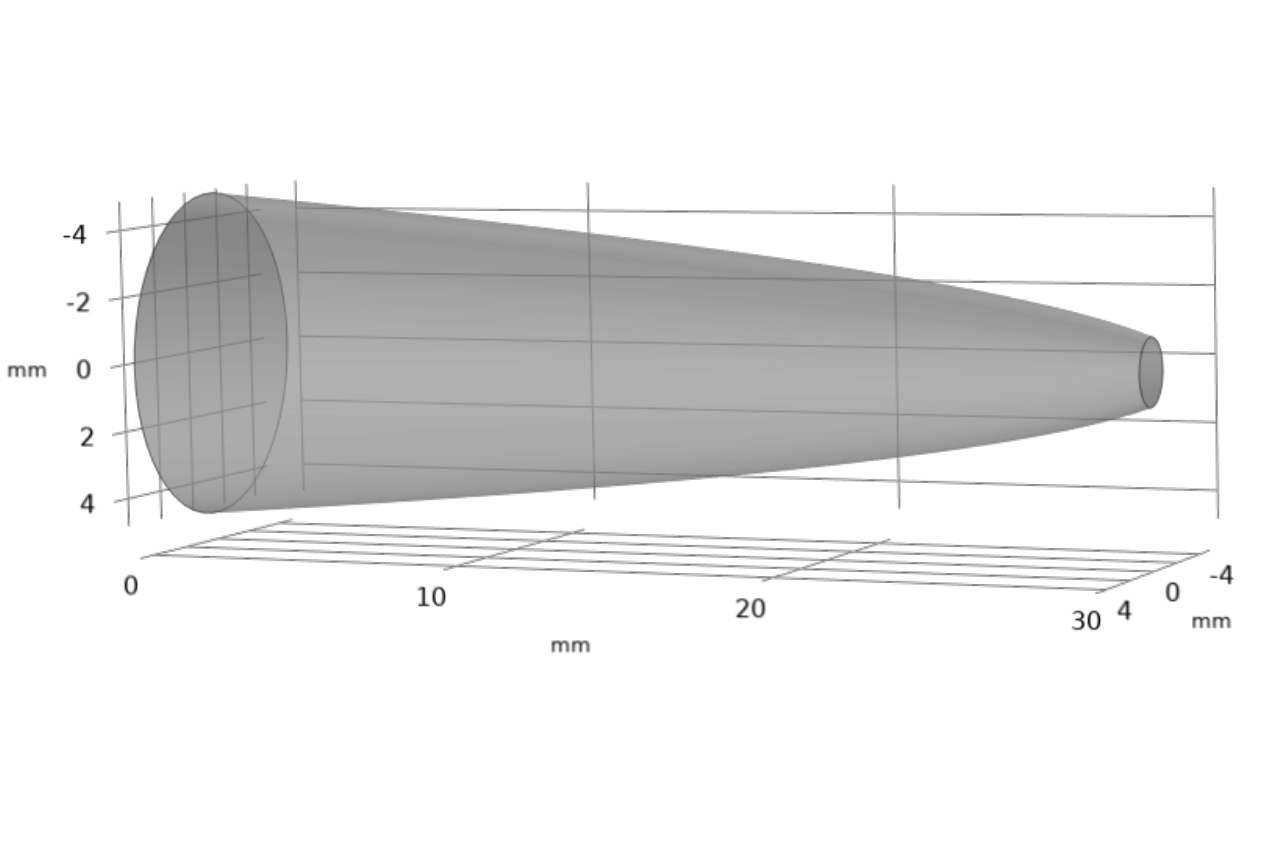}
				\includegraphics[trim={1.2cm 3.5cm 0cm 2.1cm},clip,width=.5\linewidth]{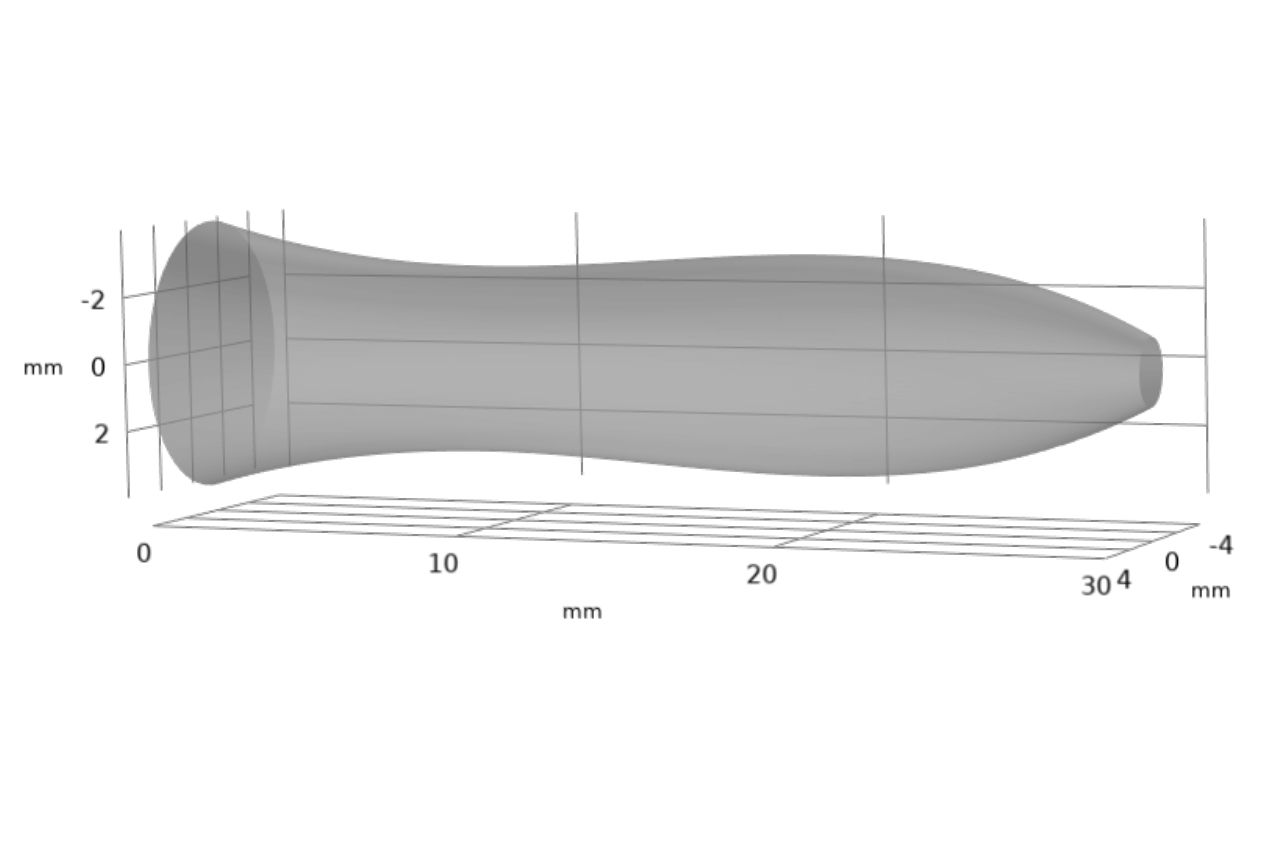}}\hfill}
		
		\caption{Validation geometries: three-dimensional geometry of the ear canal simulator as used in \citet{rasetshwane2011inverse} (top), a parabolic horn and a tapered parabolic horn (bottom).}
		\label{fig:ECsimulator_mesh_geo}
		\vskip1in
	\end{figure}

\begin{figure}[H]
		\centering
		\baselineskip=12pt

		\includegraphics[trim={0cm 0.5cm 0cm 0cm},clip,width=0.5\linewidth]{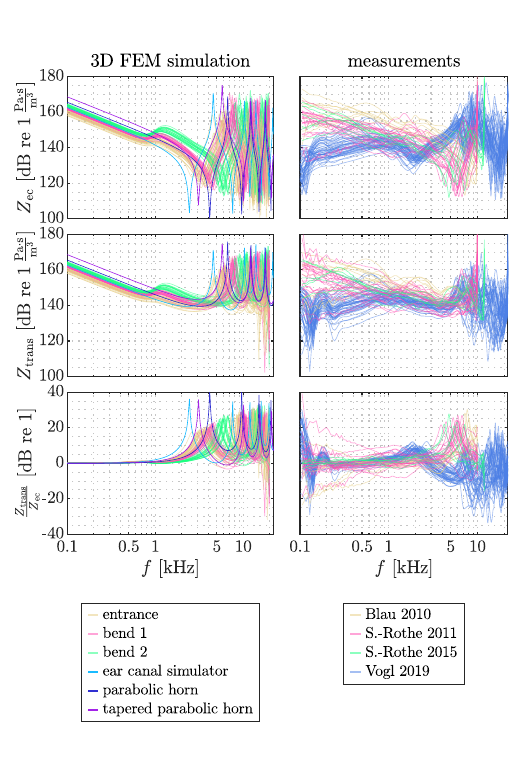}

		\caption{Level of $Z\rm{_{ec}}$ (top), $Z\rm{_{trans}}$ (middle) and $Z\rm{_{trans}}$/$Z\rm{_{ec}}$ (bottom), derived by three-dimensional FEM simulations on 21 ear canals, each cut at the entrance and at the first and second bends, as well as on artificial, ear canal-like geometries (left column) and derived from 84 measurements on ear canals from four different databases (right column).}

		\label{fig:Zin_Ztrans}
	\end{figure}

	\textbf{Simulations using ear canal simulator and parabolic horn geometries}\\
	To reproduce the results of \citet{rasetshwane2011inverse}, a three-dimensional-revolved FEM simulation for the ear canal simulator introduced in \citet{rasetshwane2011inverse} was carried out. For this purpose, the geometry shown in Fig.~\ref{fig:ECsimulator_mesh_geo} at the top was taken from Fig.~2 of \citet{rasetshwane2011inverse} as accurately as possible. In order to validate the one-dimensional electro-acoustic model in Section~\ref{sec:EAmod} with respect to three-dimensional FEM simulations, the input and transfer impedances of a simple parabolically shaped horn and a tapered variant that resembles an ear canal were also simulated, see Fig.~\ref{fig:ECsimulator_mesh_geo} (bottom). All surfaces were considered to be rigid with no further boundary conditions except the normal velocity excitation at the larger terminating surface on which the input impedance $Z\rm{_{ec}}$ was determined. The transfer impedance $Z\rm{_{trans}}$ was calculated using the pressure on the center axis at 3.5\,mm before the terminating surface. This choice is motivated by the observation that, on average, the projection of the umbo point on the CL for the present ear canal geometries is 3.5\,mm in front of the innermost corner.\\

	\subsubsection{\label{sec:meas} Measurements of input and transfer impedances}
	In order to validate the inverse solution with measurement data, four studies where input and transfer impedances of (residual) ear canals on human subjects were determined were considered \cite{blau2010prediction} \cite{sankowsky2011prediction} \cite{sankowsky2015individual} \cite{vogl2019individualized}. In these studies, the input sound pressure $p\rm{_{mic}}$ on which $Z\rm{_{ec}}$ was based was measured at different insertion depths, depending on the used probe. Either individual or generic ear molds, which partially had a vent, were used.
	In addition, the sound pressure at the eardrum $p\rm{_d}$ was measured in each of the studies, using a probe tube microphone so that the transfer impedance $Z\rm{_{trans}}$ of the ear canals could also be determined. The measurement of $p\rm{_d}$ was carried out either simultaneously with or sequentially after the measurement of the input sound pressure $p\rm{_{mic}}$. Table \ref{tab:databases} lists the most important details of the studies.
	14 of 20 data samples had to be removed from \citet{sankowsky2015individual} due to unexpected phase curves which, according to the authors, were not correctly modeled in their post-processing, cf. Fig.~5 in \citet{sankowsky2015individual}. The remaining samples show the typical phase shift at the first characteristic minimum of $Z\rm{_{ec}}$ (not shown here).\\
	Measured data from all studies often show flattened $Z\rm{_{ec}}$ levels at low frequencies in comparison to simulated data. This can be explained by leakage around the impedance probe and vent effects not included in the model determining $Z\rm{_{ec}}$ from the probe measurement. This effect appeared to be particularly large in \citet{vogl2019individualized}, since in that study the probe tube for the simultaneous measurement of $p\rm{_d}$ bypassed the ear mold. The level difference between transfer and input impedances at low frequencies should equal 0\,dB, but deviates in parts significantly. Also, the placement of the probe in the successive measurement of $p\rm{_{mic}}$ and $p\rm{_d}$ appeared to have introduced additional errors in measurements in \citet{blau2010prediction} and \citet{sankowsky2011prediction}. Nonetheless, most measurements are comparable to simulations. No further data entries were removed.\\
	Measurement data in the databases have different $f\rm{_{lim}}$ at 10, 12 or 22.05\,kHz, as can be seen in the Table~\ref{tab:databases}. Data from \citet{blau2010prediction} and \citet{sankowsky2011prediction} are available between 0.1\.kHz and 10\,kHz, data from \citet{sankowsky2015individual} for the frequency range between 0.1\,kHz and 12\,kHz and data from \citet{vogl2019individualized} are available between 0 and 22.05\,kHz.\\

\begin{table}[b]
	
	\footnotesize

	\begin{rotatetable}	
		\centering
		\caption{Details on measurement databases}
		\label{tab:databases} 
			\centering
			\begin{tabular}{c| l | l | l | l }
				
				\hline\hline
				database &\citet{blau2010prediction}& \citet{sankowsky2011prediction} & \citet{sankowsky2015individual}  & \citet{vogl2019individualized}      \\
				\hline\hline
				N number of ears   & N=10   (cadaver ears)      & N=20 (pathologically normal)                     & N=6 (without phase errors)         & N=24 (48 measurements for 2 ear molds)           \\
				\hline
				ear mold                 & individual,    closed      & individual,   closed,                  & individual, vented, closed in 4 cases & individual and generic,   vented \\
				\hline
				insertion depth          & beyond 2nd bend,                  & $\approx$ 2nd bend,                              & between 1st and 2nd   bend (itc)      & shortly before 1st   bend        \\
				\hline
				measurement   $p\rm{_d}$,   $p\rm{_{mic}}$ & successive                 & successive                             & quasi   simultaneous                  & simultaneous                     \\
				\hline
				frequency range          & 0.1~–~10\,kHz           & 0.1~–~10\,kHz                         & 0.1~–~12\,kHz                        & 0~–~22.05\,kHz          \\
				\hline\hline
			\end{tabular}
	\end{rotatetable}
	
\end{table}
	\newpage

	\subsection{\label{sec:present_procedure} Modified method to estimate $Z\rm{_{trans}}$}
	
	The review in Section~\ref{sec:procedure_of_previous_work} motivates a more thorough investigation of parameters and further adjustments of the method to obtain the inverse solution of Webster's horn equation. The following steps of processing were finally included:\vspace{+0.15cm}\\ 
	\textbf{I. Extra-/Interpolation:} After determining $Z\rm{_{ec}}$ for a limited frequency range up to $f\rm{_{lim}}$, missing values at frequencies up to $f\rm{_{sup}}/2$ were extrapolated. Many methods of extrapolations were conceivable, such as a constant at zero, at a mean value across all frequencies or at the last value at $f\rm{_{lim}}$. We finally opted for extrapolating $Z\rm{_{ec}}$ by zero-padding, i.e. $Z\rm{_{ec}}$\,=\,0 for $f\rm{_{lim}}$\,$<$\,$f$\,$<$\,$f\rm{_s}/2$, (without any amplitude correction). This implied an extrapolated frequency domain reflectance to minus one.\\
	If values at low frequencies are missing, it is also necessary to extrapolate to low frequencies. Due to the limitations of microphones and loudspeakers, this often affects frequencies below 100\,Hz in measurements.
	In preliminary considerations, the method appeared to be robust against extrapolation of $Z\rm{_{ec}}$\ to less accurate values at low frequencies. A constant at the first valid value of $Z\rm{_{ec}}$ was used in the present study.\\
	For a subsequent interpolation of the entire frequency range (between 0\,Hz and $f\rm{_{sup}}$), the FFT length was set to ensure a frequency resolution of at least 80\,Hz, with a minimum of 2$^{12}$.\vspace{+0.15cm}\\ 
	\textbf{II. Windowing $R(x,\omega)$:} The frequency domain reflectance $R(x,\omega)$ at $x$\,=\,0 was generated by Eq.~(\ref{eq:R}) with a first guess of $Z\rm{_0}(x=0)$ and the Blackman window low pass filtering was applied. Contrary to the method in \citet{rasetshwane2011inverse}, the cut-off frequency $f\rm{_{cut}}$ may also be above $f\rm{_s/2}$ which made the previous extrapolation come into effect, since extrapolated values of the frequency domain reflectance were then not completely suppressed. Values greater than $f\rm{_{cut}}$ up to $f\rm{_{sup}}/2$ were set to zero.\vspace{+0.15cm}\\ 
	\textbf{III. Adjusting the characteristic impedance and $R(x,\omega)$:} The characteristic impedance at the entrance $Z\rm{_0}(x=0)$, and therefore also $R(x,\omega)$, was adjusted as described in Section~\ref{sec:procedure_of_previous_work} based on \citet{rasetshwaneSourceCode2012old} ('surge~I') or based on \citet{rasetshwaneSourceCode2012} ('surge~II').\vspace{+0.15cm}\\ 
	\textbf{IV. Inverse solution $A(x)$ from TDR:} The updated and Blackman windowed $R(x,\omega)$ at $x$=0 was transformed to the time domain (without time-reversed addition). This time domain reflectanc was then used as input to calculate the inverse solution $A(x)$ of Webster's horn equation via the finite difference approximation (as addressed in \ref{sec:inversesolution} and taken from \citet{rasetshwaneSourceCode2012old}) up to $x$\,=\,$l_{max}$\,=\,50\,mm.\vspace{+0.15cm}\\ 
	\textbf{V. Transfer impedance:} Finally, the inverse solution $A(0\le x\le l)$ is resampled to a common spatial resolution $\Delta x$\,=\,0.1\,mm and fed to the EA model for different lengths $l$, to eventually calculate $Z\rm{_{trans}}$.\vspace{+0.15cm}\\
	In the following, this method is referred to as the method of the present study.\\
	\newpage
	\section{\label{sec:results} Results}
	Following the reproduction of results for the ear canal simulator as utilized in \citet{rasetshwane2011inverse} and the validation of the EA model, the method of the present study described in Section~\ref{sec:present_procedure} was \textit{calibrated} by investigating systematically the most advantageous parameters, so that the EA model fed by the inverse solution would provide the most accurate $Z\rm{_{trans,mod}}$ compared to $Z\rm{_{trans,ref}}$ obtained from three-dimensional FEM simulations. Lastly, the calibrated method was applied to measurements to \textit{validate} the findings.\\
	
	\subsubsection{\label{sec:review} Reproduction of the inverse solution of Webster's horn equation for an ear canal simulator}
\begin{figure}[H]
		\centering
		
		\figline{
			
			\includegraphics[trim={0cm 0.7cm 0cm 0.2cm},clip,width=0.5\linewidth]{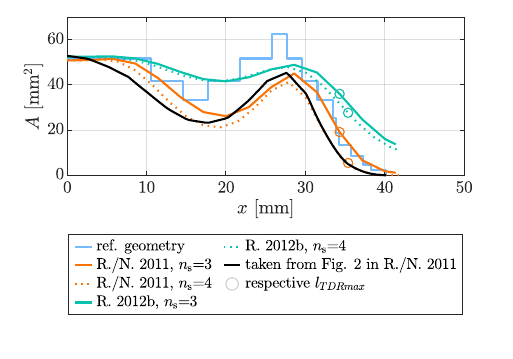}	
			
		}
		
	\caption{\label{fig:RasNeelyfig2}{Reference geometry of the ear canal simulator and results for measured data taken from Fig. 2 in \citet{rasetshwane2011inverse} and results applying the method by \citet{rasetshwane2011inverse} (conforming with \citet{rasetshwaneSourceCode2012old}) and the modified method by \citet{rasetshwaneSourceCode2012} to three-dimensional FEM simulated data for the ear canal simulator (calculated area functions ending at $l\rm{_{TDR50}}$ for $f\rm{_{s}}$=40\,kHz, $f\rm{_{cut}}$\,=\,17\,kHz and a $n\rm{_{s}}$ of 3 and 4, where $f\rm{_{lim}}$\,=\,$f\rm{_{s}}$/2, $f\rm{_{sup}}$\,=\,$n\rm{_{s}}$$\cdot$$f\rm{_{s}}$).}}
		\label{fig:neely2}
	\end{figure}

	A three-dimensional FEM simulation for the ear canal simulator with $f\rm{_s}$\,=\,40\,kHz, respectively $f\rm{_{lim}}$\,=\,20\,kHz, interpolated to match a FFT length of 2$^{12}$ was carried out for the present study (see Section~\ref{sec:sim}), to reproduce results of \citet{rasetshwane2011inverse} by the use of MATLAB scripts provided as supplementary material \cite{rasetshwaneSourceCode2012old}. Applying the same method, as outlined in Section~\ref{sec:procedure_of_previous_work}, with $f\rm{_{cut}}$\,=\,17\,kHz but an upsampling factor $n\rm{_{s}}$ of 3 (not 4), produced the most similar match to the true area function, shown in Fig.~\ref{fig:neely2}. Note that a similar underestimation of the area function as reported by \citet{rasetshwane2011inverse} was observed using the measured input impedance of the ear canal simulator sampled at $f\rm{_s}$\,=\,48\,kHz using an upsampling factor $n\rm{_{s}}$ of 4 (cf. Fig.~2 in \citet{rasetshwane2011inverse}).
The difference between the original results (black solid line in Fig.~\ref{fig:neely2}) and the reproduction of present study is presumably due to small differences in the reference geometry (taken from the diagram in Fig. 2 of \citet{rasetshwane2011inverse}, and to deviations between measured \cite{rasetshwane2011inverse} and FEM simulated (present study) impedances.\\ 
For different parameter choices for $f\rm{_{cut}}$ and $f\rm{_{sup}}$, the inverse solution varied widely. This was also the case for a modified $f\rm{_s}$\, as the data could not be reproduced exactly for $f\rm{_s}$\,=\,40\,kHz, demonstrating the need for careful parameter selection and the potential to obtain a more accurate area function by adjusting the parameters. The impact of the divergent approaches in \citet{rasetshwaneSourceCode2012old} compared to \citet{rasetshwaneSourceCode2012} is further illustrated in Fig.~\ref{fig:neely2} for the case of the ear canal simulator calculated for $n\rm{_{s}}$\,=\,3 and $f\rm{_{cut}}$\,=\,17\,kHz. The effect of the time-reversed addition (removed in \citet{rasetshwaneSourceCode2012}) appeared negligible. The algorithms 'surge I' and 'surge II' adjusting $Z\rm{_0}(x=0)$ yielded similar results. On the other hand, the amplitude correction (also removed in \citet{rasetshwaneSourceCode2012}) changed the inverse solution considerably.\\
	
	\subsubsection{\label{sec:EAmod_validation} Validation of the electro-acoustic model}
\begin{figure}[H]
		\baselineskip=12pt
		\centering
		
		\includegraphics[trim={0cm 0.25cm 0cm 0cm},width=0.5\linewidth]{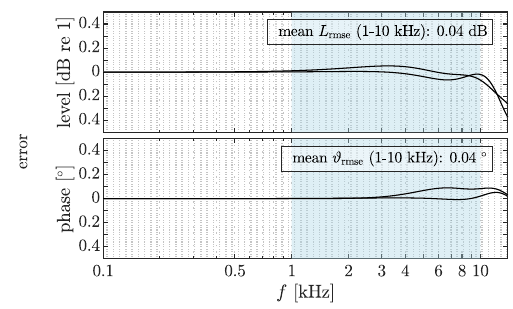}
		
		\caption{Validation of one-dimensional EA model: level and phase differences between transfer impedances calculated by one-dimensional EA model re simulated by three-dimensional FEM model for the two horns shown at the bottom of Fig.~\ref{fig:ECsimulator_mesh_geo}, using a spatial resolution $\Delta x$ of 0.1\,mm; the frequency range of interest (1\,-\,10\,kHz) is shaded blue.}

	\label{fig:errreals2ea}
\end{figure}

	For validation purposes, $Z\rm{_{trans,ref}}$ was determined for the parabolic horns shown in the lower illustration of Fig.~\ref{fig:ECsimulator_mesh_geo}. The known area functions with spatial resolutions $\Delta x$ of 1\,mm and 0.1\,mm were fed into the EA model as described in Section~\ref{sec:EAmod} to calculate transfer impedances $Z\rm{_{trans,mod}}$. Fig. \ref{fig:errreals2ea} shows the level and the phase of the error between one-dimensional and three-dimensional simulations. The mean of the root mean squared (rms) level and phase errors, ${L}\rm{_{rmse}}$ and ${\vartheta}\rm{_{rmse}}$, calculated by

	\begin{eqnarray}
		{L}\rm{_{rmse}}=\sqrt{    \frac{\sum _{\textit{f}=0.1}^{10\,\rm{kHz}}\left(20\, lg\left(\left|\frac{{\textit{Z}}\rm{_{trans,mod}}}{{\textit{Z}}\rm{_{trans,ref}}}\right|\right)\right)^2 }{\textit{n$_f$}}   }~\rm{dB}
		\label{eq:Lrmse}
	\end{eqnarray}

	and 
	\begin{eqnarray}
		{\vartheta}\rm{_{rmse}}=\sqrt{\frac{\sum _{\textit{f}=0.1}^{\rm{10\,kHz}}\left(arg\left(\frac{{\textit{Z}}\rm{_{trans,mod}}}{{\textit{Z}}\rm{_{trans,ref}}}\right)\right)^2}{\textit{n$_f$}}} \cdot \frac{180^\circ}{\pi}
		\label{eq:Prmse}
	\end{eqnarray}
	with $n\rm{_f}$ being the number of frequencies included in the frequency range between 1\,-\,10\,kHz and distributed linearly in steps of 100\,Hz, are also shown. This frequency range was chosen since the context of the present work pertains to speech applications up to 10\,kHz and since $Z\rm{_{trans}}$ equals $Z\rm{_{ec}}$ below 1\,kHz as described in Section~\ref{sec:EAmod}. Level errors did not exceed $\pm$0.06\,dB between 1 and 10\,kHz with a mean rms level error of 0.04\,dB. The absolute phase error was below 0.11\,$^\circ$ between 1 and 10\,kHz with a mean rms phase error of 0.04\,$^\circ$.
	For $\Delta$$x$\,=\,1\,mm, level errors increased up to $\pm$\,0.5\,dB, with a mean rms level error of 0.23\,dB. The absolute phase error was below 0.15\,$^\circ$ with a mean rms phase error of 0.04\,$^\circ$.\\
	A resolution of 0.1\,mm can be considered to be sufficiently accurate and is used in the following.

\subsection{\label{sec:results_calibration} Method calibration}
\subsubsection{\label{sec:lme} Transfer impedance error as a function of the termination length of the inverse solution}
\begin{figure}[H]
	\centering
	\includegraphics[trim={0cm 0.25cm 0cm 0cm},width=0.5\linewidth]{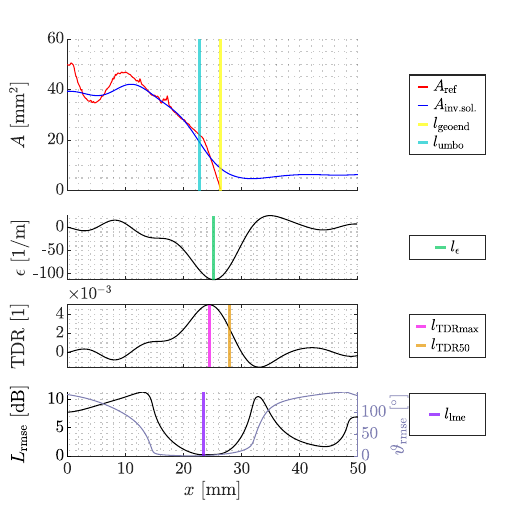}
	\caption{\label{fig:qQ_ec}{Exemplary data of the inverse solution for one ear cut at the entrance plane. From top to bottom: estimated $A\rm{_{inv.sol.}}(\textit{x})$ and reference $A\rm{_{ref}}(\textit{x})$ area functions; gradient of the logarithmic area function $\epsilon(x)$ of the inverse solution; time domain reflectance TDR$(x)$; mean (across frequency) rms errors of the transfer impedance level (Eq. (\ref{eq:Lrmse})) and phase (Eq. (\ref{eq:Prmse})). In the bottom diagram, the optimal termination length $l\rm{_{lme}}$ defined at the minimum of $L\rm{_{rmse}}(\textit{x})$ is also shown. The definitions of the other lenghts are given in Section~\ref{sec:termination}.}}
	\label{fig:example}
	
\end{figure}

At the top of Fig.~\ref{fig:example}, the inverse solution is shown for one exemplary ear canal cut at the first bend. Also shown are the reference area function known from extraction with the vmtk-toolbox, and different lengths to be introduced in the next section. The optimal length for the inverse solution at which the resulting transfer impedance error is minimal is of interest. For an inverse solution calculated to a maximum length $l\rm{_{max}}$, each $x$ in 0$<$$x$$\le$$l\rm{_{max}}$ was considered to be a possible termination length of the inverse solution. All these possible solutions were passed to the one-dimensional EA model to calculate $Z\rm{_{trans,mod}}$ for each step. Then, each $Z\rm{_{trans,mod}}$ was compared to the simulated reference $Z\rm{_{trans,ref}}$ so that $L\rm{_{rmse}}$ (Eq.~(\ref{eq:Lrmse})) and $\vartheta\rm{_{rmse}}$ (Eq.~(\ref{eq:Prmse})) for frequencies $f$ between 1\,kHz and 10\,kHz could be determined for each iteration as a function of $x$, shown at the bottom of Fig.~\ref{fig:example}. The length at the least mean level error of the transfer impedance associated with the global minimum of $L\rm{_{rmse}}(\textit{x})$ is termed $l\rm{_{lme}}$. $l\rm{_{lme}}$ is an intermediate variable that can only be determined using reference data and is not known in the use case. In following sections, it will serve as a basis for determining the optimal parameters that can then be used for the use case without any reference data. Note that the length associated with the least mean level
error $L\rm{_{lme}}$ does not necessarily coincide with that of the least
mean phase error $\vartheta\rm{_{lme}}$. However, the mean phase error $\vartheta\rm{_{rmse}}(\textit{x})$ is reasonably small over a wide range of $x$. Therefore, the length associated with the least mean level error $L\rm{_{lme}}$ is regarded as the optimal termination length and used in the following.

\subsubsection{\label{sec:results_delta_x} Reflectance upsampling for a sufficient spatial resolution}
\begin{figure}[H]
	\centering
	\includegraphics[trim={0cm 0.25cm 0cm 0.25cm},width=0.5\linewidth]{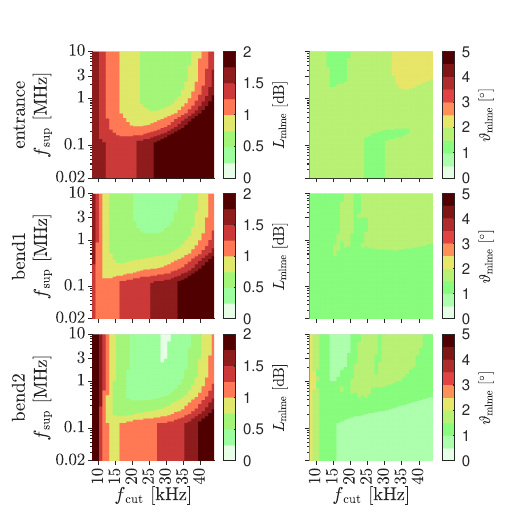}
	
	\caption{Dependency of the mean value of $L\rm{_{lme}}$ across all 21 ears ($L\rm{_{mlme}}$) and of the mean value of $\vartheta\rm{_{lme}}$ across all 21 ears ($\vartheta\rm{_{mlme}}$) on $f\rm{_{sup}}$ and $f\rm{_{cut}}$ for ear canals cut at the entrance, first and second bend (from top to bottom) at $f\rm{_{lim}}$\,=\,20\,kHz.}
	\label{fig:surf1}
	\vskip1in
\end{figure}

Keeping $f\rm{_{lim}}$ constant at 20\,kHz for different $f\rm{_{cut}}$=[8,9...,44]\,kHz and $f\rm{_{sup}}$=[0.02,0.04,...0.12, 0.125,0.144,0.16,0.192, 0.25,0.5,...2,2.5,...5,7.5,10]\,MHz, the least mean level errors $L\rm{_{lme}}$ and phase errors $\vartheta\rm{_{lme}}$ were determined for all 21 ear canal geometries cut at 3 different lengths, for each combination of $f\rm{_{cut}}$ and $f\rm{_{sup}}$, employing the method of the present study (with 'surge I').
The mean of all 21 $L\rm{_{lme}}$- and $\vartheta\rm{_{lme}}$-values for one group of cut planes is shown as the $L\rm{_{mlme}}$ and $\vartheta\rm{_{mlme}}$ in Fig.~\ref{fig:surf1}. One can observe that $L\rm{_{mlme}}$ decrease for medium and higher $f\rm{_{sup}}$. A saturation effect can be observed with only minor improvement above $f\rm{_{sup}}$\,=\,3.5\,MHz. Above $f\rm{_{sup}}$\,=\,3.5\,MHz, a $L\rm{_{mlme}}\,<\,$0.6\,dB is reached for $f\rm{_{cut}}$ between 20\,kHz and 35\,kHz for all three groups of cut planes. The shorter the ear canals, the smaller the minimal $L\rm{_{mlme}}$. The mean phase error $\vartheta\rm{_{mlme}}$ slightly increases in the field of lowest level errors (visible for group 'bend 2'), but is still limited to less than 2\,$^\circ$. In the following, $f\rm{_{sup}}$\,=\,3.5\,MHz is considered as a sufficiently high spatial resolution with $\Delta x$\,$\approx$\,0.1\,mm for $c$\,=\,351.8\,$\frac{m}{s}$ (for a temperature in the ear canal of $T$\,=\,308\,K).

\subsubsection{\label{sec:results_adjustment_of_Z0} Adjustment of $Z\rm{_0}(\textit{x}=0)$}
\begin{figure}[H]
	\centering
	\includegraphics[trim={0cm 0.25cm 0cm 0cm},width=0.5\linewidth]{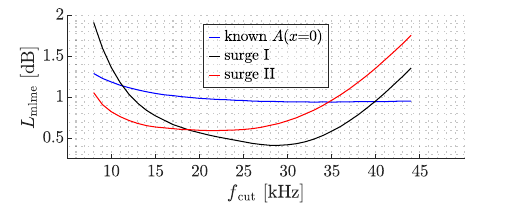}
	
	\caption{Dependency of $L\rm{_{mlme}}$ on $f\rm{_{cut}}$ for different methods adjusting the characteristic impedance, for all ear canals with $f\rm{_{lim}}$\,=\,20\,kHz and $f\rm{_{sup}}$\,=\,3.5\,MHz.}
	\label{fig:surge}
	\vskip1in
\end{figure}

In Section~\ref{sec:procedure_of_previous_work}, two approaches ('surge I' and 'surge II') were presented for adjusting the characteristic impedance. A small difference between these two approaches was observed in the resulting area funtions for the ear canal simulator in Section~\ref{sec:review}. In order to analyze the resulting accuracy of transfer impedance predictions,
the least mean level error $L\rm{_{lme}}$ was determined for all 21 ear canal geometries cut at the three different lengths, using $f\rm{_{lim}}$\,=\,20\,kHz and $f\rm{_{sup}}$\,=\,3.5\,MHz, at different $f\rm{_{cut}}$=[8,9...,44]\,kHz. The mean of all $L\rm{_{lme}}$ values $L\rm{_{mlme}}$ is again used as error indicator, and is shown as a function of $f\rm{_{cut}}$ in Fig.~\ref{fig:surge}. Given the availability of a geometric reference, a third method utilizes the actual value of $A(x=0)$ in Eq. (\ref{eq:Z0}) to determine $Z_0(x=0)$. As can be seen in Fig.~\ref{fig:surge}, the optimal performance can be achieved utilizing 'surge I' at $f\rm{_{cut}}$\,=\,28\,kHz.

\subsubsection{\label{subsubsec:5:1:3} Linear model for optimal $f\rm{_{cut}}$ re $f\rm{_{lim}}$}
\begin{figure}[H]
	\centering
	\includegraphics[trim={0cm 0.25cm 0cm 1.2cm},width=0.5\linewidth]{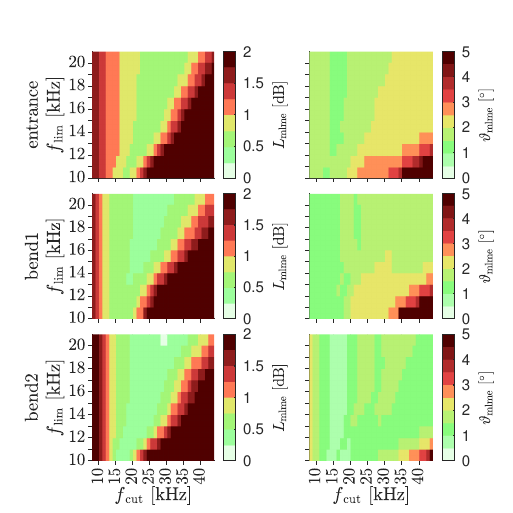}
	
	\caption{Dependency of the mean value of $L\rm{_{lme}}$ across all 21 ears ($L\rm{_{mlme}}$) and of the mean value of $\vartheta\rm{_{lme}}$ across all 21 ears ($\vartheta\rm{_{mlme}}$) on $f\rm{_{lim}}$ and $f\rm{_{cut}}$ for ear canals cut at the entrance, first and second bend (from top to bottom) at $f\rm{_{sup}}$\,=\,3.5\,MHz.}
	\label{fig:surf2}
\end{figure}

\begin{figure}[H]
	\centering
	 \includegraphics[trim={0cm 0.25cm 0cm 0cm},width=0.5\linewidth]{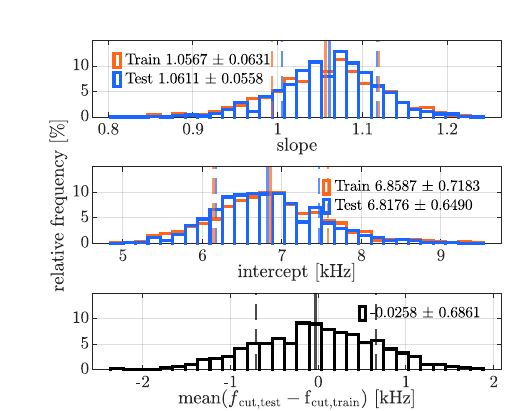}
	
	\caption{Distributions of the estimated slopes, intercepts, and the mean group error differences between tested and trained $f\rm{_{cut}}$ from repeated cross-validation (1000 iterations, train/test split: 10/11 ear canals cut at three different lengths).}
	\label{fig:lin_reg_cross}
\end{figure}

\begin{figure}[H]
	\centering
 \includegraphics[trim={0cm 0.25cm 0cm 0cm},width=0.5\linewidth]{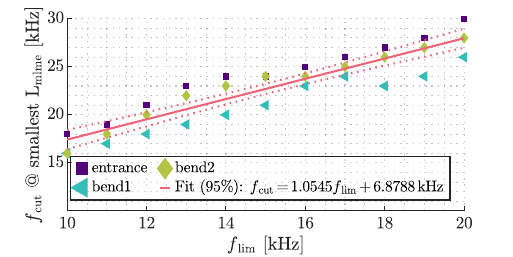}
	
	\caption{Linear regression model for the relation between $f\rm{_{lim}}$ and $f\rm{_{cut}}$ at minima of $L\rm{_{mlme}}$, extracted from data shown in Fig.~\ref{fig:surf2}.}
	\label{fig:lin_reg}
\end{figure}

Based on the results from Sections~\ref{sec:results_delta_x} and \ref{sec:results_adjustment_of_Z0}, 'surge I' is applied and $f\rm{_{sup}}$ is kept constant at 3.5\,MHz in the following. The dependency of $L\rm{_{mlme}}$ and $\vartheta\rm{_{mlme}}$ on the parameters $f\rm{_{lim}}$ and $f\rm{_{cut}}$ is shown in Fig.~\ref{fig:surf2}, separately for the groups of cut planes. The higher $f\rm{_{lim}}$, the wider the interval of advantageous $f\rm{_{cut}}$. $\vartheta\rm{_{mlme}}$ essentially does not exceed 2\,$^\circ$ where $L\rm{_{mlme}}$ is small and is therefore not considered in the following.\\
In Fig.~\ref{fig:lin_reg}, values of $f\rm{_{cut}}$ at minimum $L\rm{_{mlme}}$ are shown as a function of $f\rm{_{lim}}$. From these data, one linear regression model for the best value for $f\rm{_{cut}}$ for a given value of $f\rm{_{lim}}$ was estimated.\\
To assess the robustness and generalizability of this linear regression model, a repeated cross-validation approach was employed \cite{hastie2009elements}. Instead of modeling each condition separately, the analysis focused on the average response across all three conditions, aiming to capture a general trend while enhancing model stability. In each of 1000 cross-validation iterations, ear canals from 10 subjects were randomly selected for model training, and the remaining 11 were excluded from all three groups of ear canal lengths and served as a test set. For both training and test subsets, group mean values were determined for each $f\rm{_{lim}}$-value. A linear regression model was then fitted to the training group mean values. Resulting model parameters - slope and intercept - were applied to the test mean values. The prediction accuracy was quantified using the error difference between predicted and actual test values.
As shown in Fig.~\ref{fig:lin_reg_cross}, the validation demonstrated a strong consistency between training and testing distributions for the slope and the intercept. The mean group difference between $f\rm{_{cut,train}}$ and $f\rm{_{cut,test}}$ is close to 0\,kHz with a standard deviation of less than 0.7\,kHz. These results indicate that the linear relationship is stable across subsamples and conditions, with limited variability in model parameters and moderate prediction error.\\
The final model used in the following was fitted to the group means for all 21 ear canals cut at the three different lengths, achieving a $R^2$ of 83\,$\%$ with 
\begin{eqnarray}
	f\rm{_{cut}}=1.05 \cdot \textit{f}\rm{_{lim}} + 6.88\,kHz\,.
	\label{eq:linMod}
\end{eqnarray}

\begin{figure}[H]
	\centering
 \includegraphics[trim={0cm 0.25cm 0cm 1cm},width=0.5\linewidth]{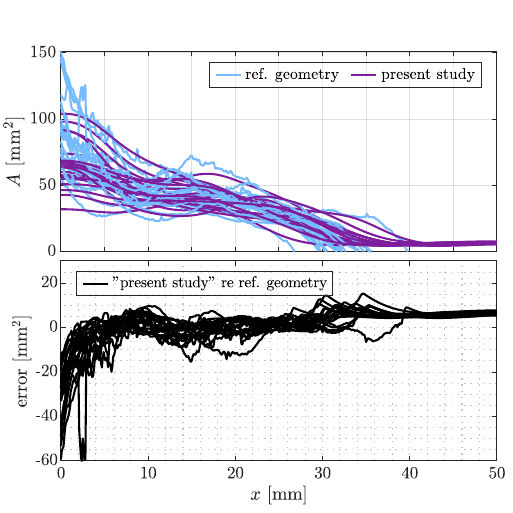}	
	\caption{Reference area functions of ear canals cut at entrance plane vs. inverse solutions as caculated in the present study for $f\rm{_{sup}}$\,=\,3.5\,MHz, $f\rm{_{lim}}$\,=\,20\,kHz and $f\rm{_{cut}}$\,=\,28\,kHz up to $x$\,=\,50\,mm.}
	
	\label{fig:COMP_VMTK_INVSOL}
\end{figure}

\begin{figure}[H]
	\centering
	\includegraphics[trim={0cm 0.8cm 0cm 0cm},width=0.5\linewidth]{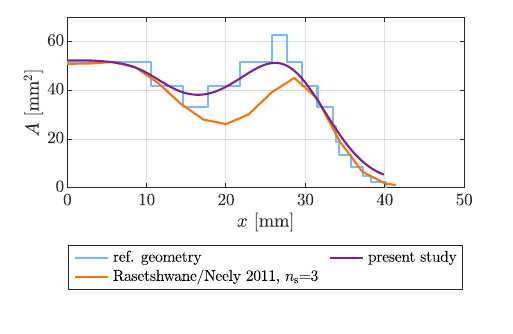}	
	
	\caption{Exact area function of the ear canal simulator and inverse solutions up to $x$\,=\,$l\rm{_{TDR50}}$ applying the method of \citet{rasetshwane2011inverse} for $f\rm{_{sup}}$\,=\,120\,kHz, $f\rm{_{lim}}$\,=\,20\,kHz and $f\rm{_{cut}}$\,=\,17\,kHz and the method modified in present study for $f\rm{_{sup}}$\,=\,3.5\,MHz, $f\rm{_{lim}}$\,=\,20\,kHz and $f\rm{_{cut}}$\,=\,28\,kHz.}
	
	\label{fig:neely3}
\end{figure}

With this regression model, an optimal $f\rm{_{cut}}$ of 28\,kHz is obtained for $f\rm{_{lim}}$\,=\,20\,kHz. Together with $f\rm{_{sup}}$\,=\,3.5\,MHz 
and 'surge I', the inverse solution of Webster's horn equation results in the area functions shown in Fig.~\ref{fig:COMP_VMTK_INVSOL} as 'present study'. 
They can be seen to agree well with the reference area functions, with the exception of a slight underestimation, especially at the lateral end, as shown in Fig.~\ref{fig:COMP_VMTK_INVSOL} for ear canals cut at the entrance. 
Also, using the same parameter choice for the ear canal simulator yields an inverse solution that better matches the true geomtry, compared to the best possible match reached with the method reported in \citet{rasetshwane2011inverse} and implemented in \citet{rasetshwaneSourceCode2012old}, see Fig.~\ref{fig:neely3}.\\

\subsubsection{\label{sec:termination} Termination length of the inverse solution}
\begin{figure}[H]
	\centering
	\includegraphics[trim={0cm 0.8cm 0cm 1.2cm},width=0.5\linewidth]{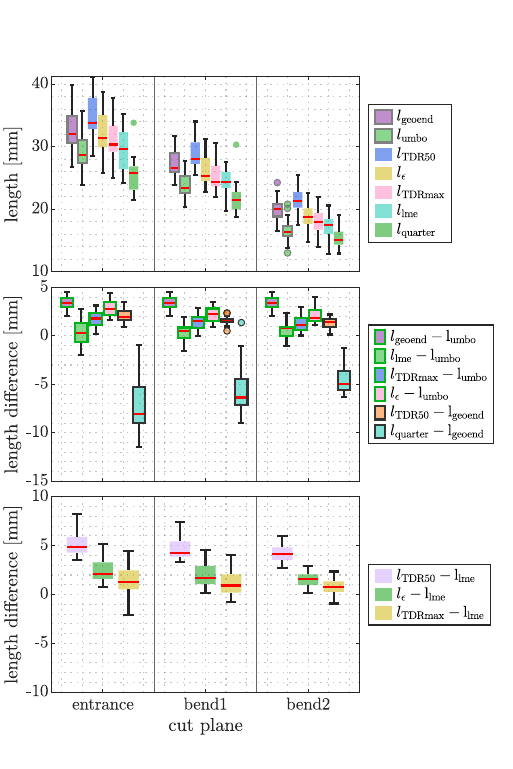}
	
	\caption{Absolute termination lengths (top), length differences (middle, bottom) from inverse solutions calculated by setting $f\rm{_{sup}}$\,=\,3.5\,MHz, $f\rm{_{lim}}$\,=\,20\,kHz, $f\rm{_{cut}}$\,=\,28\,kHz}
	\label{fig:Length_sim}
\end{figure}

For an optimal estimation of the transfer impedance, the inverse solution needs to be limited at $l\rm{_{lme}}$ (introduced in Section~\ref{sec:lme}). This value is not known in the application case. In the following, different termination lengths are examined. Fig.~\ref{fig:example} shows some lengths that are known from the geometric ear canal models or can be derived from inverse solutions. First, there is the geometric length $l\rm{_{geoend}}$, defined between the lateral end of the ear canal and the innermost point of the centerline. Second, the length $l\rm{_{umbo}}$ between the lateral end of the ear canal and the umbo point, projected on the centerline, may be of interest. Furthermore, the maximum of the time domain reflectance can be used to infer a length $l\rm{_{TDRmax}}$ which is related to the reflection at the eardrum. In \citet{rasetshwaneSourceCode2012}, the end of the ear canal is associated with $l\rm{_{TDR50}}$ where the time domain reflectance has decayed to 50$\%$ of its maximum. The global minimum of the gradient of the logarithmic area function $\epsilon$ at the length $l\rm{_{\epsilon}}$ also appears to be prominent.\\
The distributions of these termination lengths of the inverse solutions (calculated using $f\rm{_{lim}}$\,=\,20\,kHz, $f\rm{_{sup}}$\,=\,3.5\,MHz and $f\rm{_{cut}}$\,=\,28\,kHz) determined for all ear canal geometries are shown in the top diagram of Fig.~\ref{fig:Length_sim}.
In addition, the quarter-wave length $l\rm{_{quarter}}$, derived from the frequency of the first characteristic minimum of the $Z\rm{_{ec}}$, is also shown. Characteristic length differences are shown in the middle and bottom diagram of Fig.~\ref{fig:Length_sim}.\\
The interquartile range of the absolute length $l\rm{_{geoend}}$ of the ear canals cut at the entrance is approximately 30 to 35\,mm, and 26 to 29\,mm and 19 to 21\,mm for ear canal geometries cut at the first and second bends, respectively. The median distance between $l_{umbo}$ and $l_{geoend}$ is observed to be between 3 and 4\,mm. The median lengths of $l\rm{_{\epsilon}}$ and $l\rm{_{TDRmax}}$ lie between the median of $l\rm{_{umbo}}$ and $l\rm{_{geoend}}$, with the median $l\rm{_{\epsilon}}$ somewhat closer to the end. The median of $l\rm{_{TDR50}}$ is about 1 to 2.5\,mm behind the geometric end. The quarter wave lengths are largely scattered and much shorter than the median of the length $l\rm{_{geoend}}$.\\
For further investigation, the difference between the lengths $l\rm{_{TDR50}}$, $l\rm{_{TDRmax}}$ and $l\rm{_{\epsilon}}$ obtained from the inverse solution, and the location of the least mean error $l\rm{_{lme}}$ is of particular importance. The medians of these differences are all positive, indicating that the termination length tends to be overestimated. The scatter is approximately the same for all lengths.\\

\begin{figure}[H]
	\centering
	
	\includegraphics[trim={0cm 0.3cm 0cm 1.1cm},clip,width=0.5\linewidth]{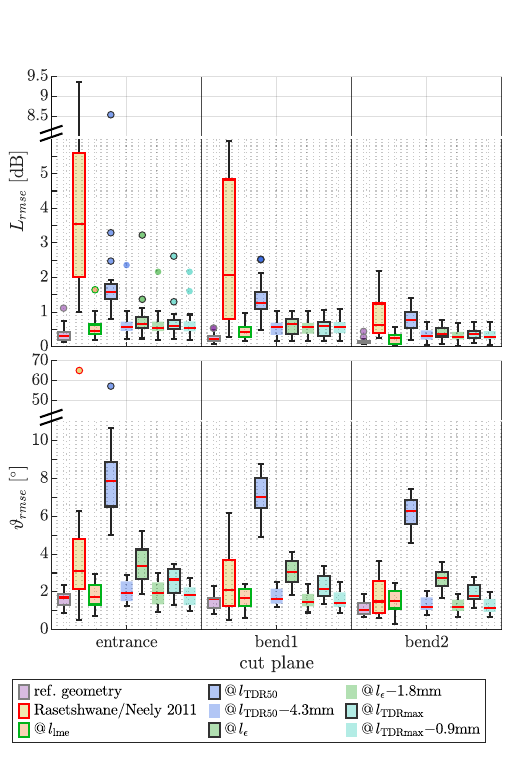}
	
	\caption{Boxplot showing values for $L\rm{_{rmse}}$ (top) and $\vartheta\rm{_{rmse}}$ (bottom), calculated for simulation-based data, using the vmtk-derived area function, applying the method by \citet{rasetshwane2011inverse} as implemented in \cite{rasetshwaneSourceCode2012old} with an optimal $n\rm{_s}$ of 3 and $f\rm{_{cut}}$\,=\,17\,kHz for $f\rm{_{lim}}$\,=\,20\,kHz, as well as applying the method of the present study at $f\rm{_{sup}}$\,=\,3.5\,MHz, $f\rm{_{lim}}$\,=\,20\,kHz, $f\rm{_{cut}}$\,=\,28\,kHz.}
	\label{fig:box_error_sim}
\end{figure}

The effect of using different termination lengths on the resulting accuracy of transfer impedance predictions is shown in Fig.~\ref{fig:box_error_sim}, together with the accuracy of the original method by \citet{rasetshwane2011inverse}. Values for $L\rm{_{rmse}}$ and $\vartheta\rm{_{rmse}}$ were determined for all ear canal geometries. Feeding the one-dimensional EA model with the vmtk-derived area function limited at the projected umbo point provides the best performance in terms of the median of $L\rm{_{rmse}}$ in comparison to the three-dimensional simulation. Using the inverse solution terminated at $l\rm{_{lme}}$ achieved slightly larger median values of less than 0.6\,dB and 2\,$^\circ$. The termination at $l\rm{_{TDR50}}$ degraded the performance more than a termination at $l\rm{_{TDRmax}}$ or $l\rm{_{\epsilon}}$, as expected, caused by the larger distance between $l\rm{_{lme}}$ and $l\rm{_{TDRmax}}$.\\
It was hypothesized that the termination of the inverse solution at distance-corrected lengths would provide a more accurate prediction of $Z\rm{_{trans}}$. The values for this length correction equal the median differences of $l\rm{_{TDR50}}$, $l\rm{_{TDRmax}}$ and $l\rm{_{\epsilon}}$ to $l\rm{_{lme}}$, derived from data of all three groups shown at the bottom of Fig.~\ref{fig:Length_sim}. With length corrections for $l\rm{_{\epsilon}}$ by 1.8\,mm and $l\rm{_{TDRmax}}$ by 0.9\,mm, there seems to be a small improvement for $L\rm{_{rmse}}$ and a prominent improvement for the corrected lengths at $l\rm{_{TDR50}}$-4.3\,mm, compared to the respective non-corrected lengths, and a general reasonable improvement concerning $\vartheta\rm{_{rmse}}$. Errors for terminating the inverse solution at corrected length produce similar scattering. Smaller median $L\rm{_{rmse}}$ can be achieved for shorter ear canals. The median $L\rm{_{rmse}}$ for the termination at corrected lengths is $\approx$\,0.15\,dB above the value reached with a termination at $l\rm{_{lme}}$. Terminating the inverse solution at $x$\,=\,$l_{ \epsilon }$\,-\,1.8\,mm yielded visually the best results. The results of the original method by \citet{rasetshwane2011inverse} could thus be noticeably improved, in particular for longer ear canals.\\

\begin{figure}[H]
	\centering
	\figcolumn{
		\includegraphics[trim={0cm 0.2cm 0cm 0.25cm},width=0.5\linewidth]{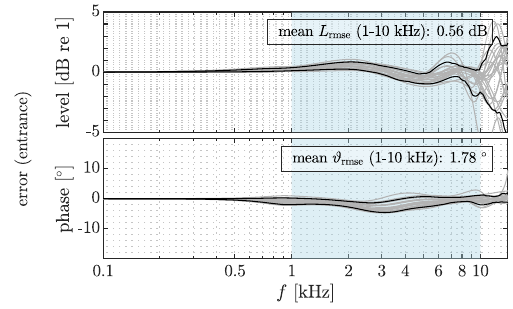}
		\includegraphics[trim={0cm 0.2cm 0cm 0.25cm},width=0.5\linewidth]{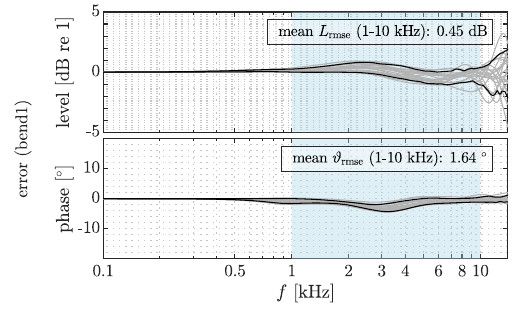}
		\includegraphics[trim={0cm 0.25cm 0cm 0.25cm},width=0.5\linewidth]{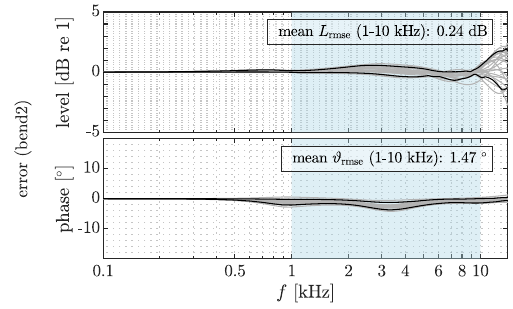}
		
	}
	
	\caption{Transfer impedance errors and 10th and 90th percentiles (with respect to the three-dimensional FEM simulation), for ear canal geometries cut at the entrance (top), at the first (mid) and the second bend (bottom): $Z\rm{_{trans,mod}}$ calculated by the inverse solution (with $f\rm{_{sup}}$\,=\,3.5\,MHz, $f\rm{_{lim}}$\,=\,20\,kHz, $f\rm{_{cut}}$\,=\,28\,kHz, 'surge I') terminated at $l\rm{_{lme}}$ and fed into one-dimensional EA model; the frequency range of interest (1\,-\,10\,kHz) is shaded blue.}
	\label{fig:error_sim_entrance}
\end{figure}

A closer examination of the frequency response of the error obtained for the inverse solution terminated at $l\rm{_{lme}}$ with $f\rm{_{sup}}$\,=\,3.5\,MHz, $f\rm{_{lim}}$\,=\,20\,kHz and $f\rm{_{cut}}$\,=\,28\,kHz reveals that the resonances are accurately captured, since this information is given by $Z\rm{_{ec}}$, but the magnitude of $Z\rm{_{trans}}$ below and between the resonances deviate, resulting in a wavy frequency dependency of the error shown in Fig.~\ref{fig:error_sim_entrance}. In addition, the errors increase above 10\,kHz as expected.\\

\subsection{\label{sec:validation} Validation using measured input and transfer impedances}
\subsubsection{\label{sec:application_lin_mod_to_meas} Application of the linear model for the selection of $f\rm{_{cut}}$ for a given $f\rm{_{lim}}$}

For given $f\rm{_{lim}}$ of the measurements presented in Section~\ref{sec:meas} from \citet{blau2010prediction}, \citet{sankowsky2011prediction}, \citet{sankowsky2015individual} and \citet{vogl2019individualized}, optimal $f\rm{_{cut}}$ values at 17.4, 17.4, 19.5 and 28\,kHz (rounded to the first decimal) were determined using the linear model from Eq.~(\ref{eq:linMod}).
In addition, values for $L\rm{_{mlme}}$ and $\vartheta\rm{_{mlme}}$ were computed as a function of $f\rm{_{cut}}$ for $f\rm{_{sup}}$\,=\,3.5\,MHz and applying 'surge I'. This analysis was limited to presumed length intervals, since otherwise $l\rm{_{lme}}$ and other termination lengths were occasionally unrealistically short or long. For example, $L\rm{_{rmse}}(\textit{x})$ showed further local minima at multiples of the ear canal length for the selected frequency range of 1 to 10\,kHz (see Fig.~\ref{fig:example}, bottom). The limits were chosen based on a length analysis for the 21 ear canal geometries shown in Fig.~\ref{fig:Length_sim} (top), yielding the following conservative assumptions: 1)~The length between entrance and umbo point does not exceed 45~mm, 2)~the length between first bend and umbo point is in the interval [15 35]~mm, and 3) the length between second bend and umbo point is in the interval [10 25]~mm. Together with the insertion depths reported by the authors of the studies containing the experimental data, conservative limits of residual ear canal lengths were derived for the given insertion depths, see Table~\ref{tab:length}.\\

\begin{table}[H]
	\centering
	\caption{Insertion depths and derived length intervals for the experimental data used for validation. Note that \citet{blau2010prediction} also reported residual ear canal lengths (from the inner face of the ear mold - beyond second bend - to the innermost point of the ear canal) between 10.8~mm and 15~mm.}
	\begin{tabular}{l|l|l}
		\hline \hline
		database & insertion depth                 & {[}min max{]} length of residual ear canal {[}mm{]}  \\
		\hline \hline
		\citet{blau2010prediction}	& beyond 2nd bend                 & {[}3 20{]}                                            \\
		\hline
		\citet{sankowsky2011prediction}		& 2nd bend                      & {[}5 30{]}                                          \\
		\hline
		\citet{sankowsky2015individual}	&	itc (between 1st and 2nd bend) & {[}10 35{]}                                        \\
		\hline
		\citet{vogl2019individualized}	&	before 1st bend                 & {[}15 45{]}              \\                          
		\hline \hline
	\end{tabular}
	\label{tab:length}
\end{table}

\begin{figure}[H]
	\centering
	\includegraphics[trim={0cm 0.25cm 0cm 0cm},width=0.5\linewidth]{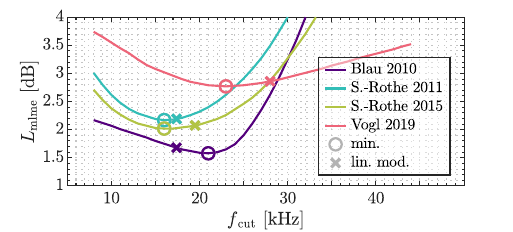}
	
	\caption{Analysis of resulting mean least mean transfer impedance level errors for measurement databases: dependency of $L\rm{_{mlme}}$ on $f\rm{_{cut}}$ for $f\rm{_{sup}}$\,=\,3.5\,MHz. Also shown are $f\rm{_{cut}}$ predicted from $f\rm{_{lim}}$ using the model of Eq.~(\ref{eq:linMod}) (crosses) and global minima of the respective measurements (circles).}
	\label{fig:MEASlinReg}
\end{figure}

Predictions for $f\rm{_{cut}}$ of the linear model were then compared to actual optimal values for $f\rm{_{cut}}$ at global minima of $L\rm{_{mlme}}$.
As for simulation-based data (see Fig.~\ref{fig:surf2}), a valley also forms for the measurements. Therefore, the model value of $f\rm{_{cut}}$ and the best possible value at the minimum can lie some distance apart without adding a major error. In fact, the additional error between the minimal possible $L\rm{_{mlme}}$ and the value at $f\rm{_{cut}}$ derived by the linear model is less than 0.1\,dB for all databases (see Fig.~\ref{fig:MEASlinReg}).

\begin{figure}[H]
	\centering
	
	\includegraphics[trim={0cm 1cm 0cm 0cm},width=0.5\linewidth]{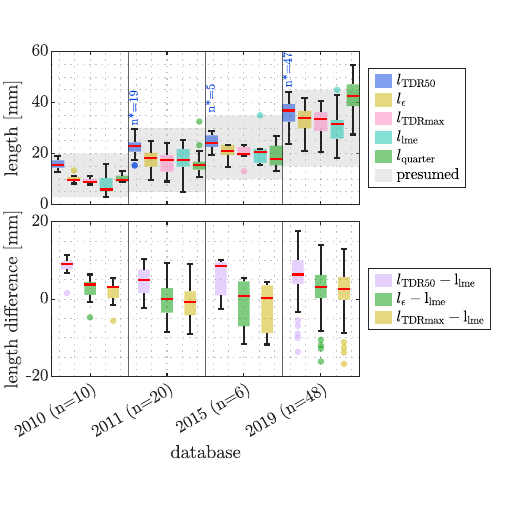}
	\caption{Results on measurements showing absolute lengths (top) and length differences (bottom), derived at $f\rm{_{sup}}$\,=\,3.5\,MHz, $f\rm{_{lim}}$ given by the respective database and optimal $f\rm{_{cut}}$ taken from the linear regression model (Eq. (\ref{eq:linMod})).}
	\label{fig:length_meas}
\end{figure}

\subsubsection{\label{subsubsec:5:2:2} Application of termination criteria to measurements}
\begin{figure}[H]
	\centering
	
	\includegraphics[trim={0cm 0.5cm 0cm 0.5cm},width=0.5\linewidth]{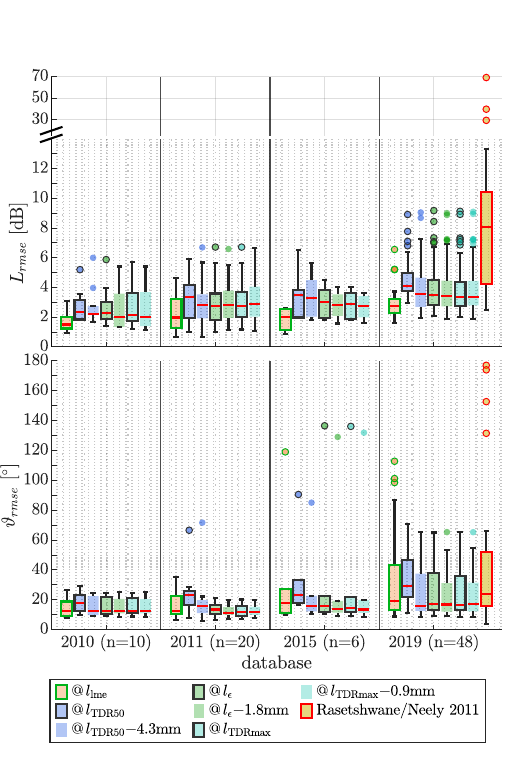}
	
	\caption{Validation results on measurements showing $L\rm{_{rmse}}$ (top) and $\vartheta\rm{_{rmse}}$ (bottom) for the method by \citet{rasetshwane2011inverse} and for the optimized method employed in the present study, with $f\rm{_{sup}}$\,=\,\,3.5\,MHz, $f\rm{_{lim}}$ given by database and optimal $f\rm{_{cut}}$ taken from the linear regression model from Eq.~(\ref{eq:linMod}).}
	\label{fig:error_meas}
\end{figure}

\begin{figure}[h]
	\centering
	\figcolumn{
		\includegraphics[trim={0cm 0.32cm 0cm 1.4cm},width=0.5\linewidth]{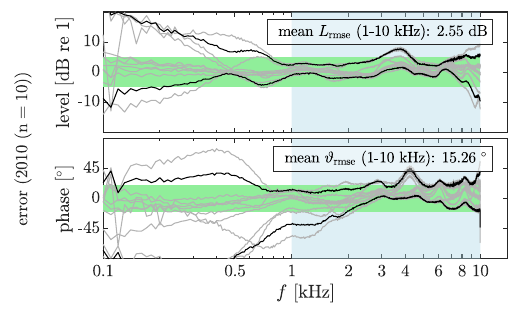}
		
		\includegraphics[trim={0cm 0.32cm 0cm 0.25cm},width=0.5\linewidth]{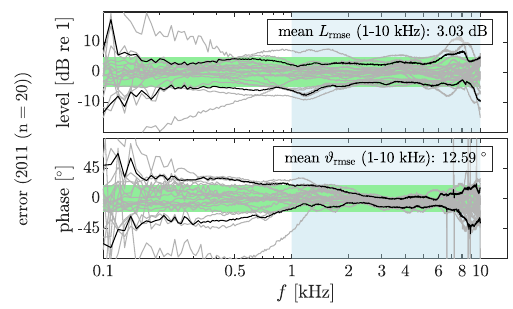}
		\includegraphics[trim={0cm 0.32cm 0cm 0.25cm},width=0.5\linewidth]{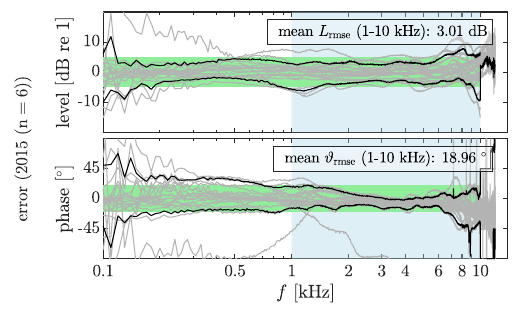}
		\includegraphics[trim={0cm 0.32cm 0cm 0.25cm},width=0.5\linewidth]{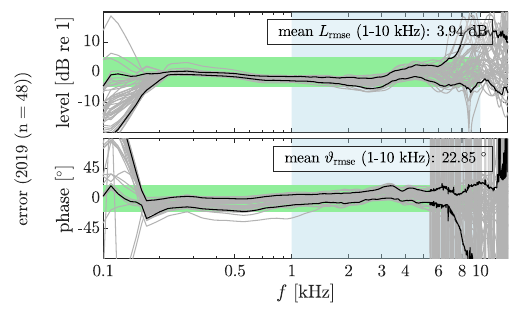}
		
	}
	\caption{Transfer impedance errors and 10th and 90th percentiles with respect to the measurements from \citet{blau2010prediction}, \citet{sankowsky2011prediction}, \citet{sankowsky2015individual}, \citet{vogl2019individualized} (from top to bottom): $Z\rm{_{trans,mod}}$ calculated by the inverse solution (with $f\rm{_{sup}}$\,=\,3.5\,MHz, database-specific $f\rm{_{lim}}$\,=\,[10 10 12 22.05]\,kHz, $f\rm{_{cut}}$\,=\,[17.4 17.4 19.5 28]\,kHz, 'surge I') terminated at $l\rm{_{\epsilon-1.8mm}}$ and fed into one-dimensional EA model; the frequency range of interest (1\,-\,10\,kHz) is shaded blue, the tolerance range is shaded green.}
	
	\label{fig:error_meas_database}
\end{figure}

In the presumed limitation intervals, with $f\rm{_{sup}}$\,=\,3.5\,MHz and $f\rm{_{cut}}$ selected using the linear model from Eq.~(\ref{eq:linMod}), derived termination lengths are shown in Fig.~\ref{fig:length_meas}.
Regarding termination length corrections, the length differences between $l\rm{_{TDR50}}$, $l\rm{_{\epsilon}}$ or $l\rm{_{TDRmax}}$ and $l\rm{_{lme}}$ reproduced the tendency of length overestimations that was already
observed for simulation-based data in Section~\ref{sec:termination}, in the same order. The length differences for measured data scatter more widely.
The time domain reflectance TDR decayed less quickly compared to simulation results. For one measurement in each of three databases, no $l\rm{_{TDR50}}$ could be identified within the presumed length interval, as indicated by the reduced sample number n* above the respective boxes.\\
Fig.~\ref{fig:error_meas} shows values of $L\rm{_{rmse}}$ and $\vartheta\rm{_{rmse}}$ derived with different termination lengths for the measurements. The median of the least mean error for three databases was slightly below 2\,dB, for one database below 3\,dB. Concerning the phase, the median of the least mean error was below 20\,$^\circ$. With an additional level error between 0.5 and 0.8\,dB and no additional phase error, the corrected and the non-corrected termination lengths $l\rm{_{\epsilon}}$ and $l\rm{_{TDRmax}}$  achieved a similar accuracy and appear to be suitable for accurate predictions.
The inter-quartile range increased slightly compared to the simulation results.\\
Regarding the original method of \citet{rasetshwane2011inverse} as implemented in \citet{rasetshwaneSourceCode2012old}, a comparison is only meaningful for the data of \citet{vogl2019individualized}, since the value of $f\rm{_{lim}}$ for this database is similar to the value used in the calibration and larger than the optimal $f\rm{_{cut}}$ of 17\,kHz for this method. Compared to results derived by the modified method employed in the present study, the accuracy of the original method is substantially lower, in particular for the level, see Fig.~\ref{fig:error_meas}.\\ 
Finally, the frequency-dependent errors of transfer impedance predictions are shown in Fig.~\ref{fig:error_meas_database} for each database. Considering measurement uncertainties, the authors defined a tolerance range of $\pm$\,5\,dB and $\pm$\,20\,$^\circ$ in the frequency interval 1\,kHz\,$\le$\,$f$\,$\le$\,10\,kHz for these databases in their respective publications. These tolerance ranges were only slightly exceeded for the present study with regard to the 10th and 90th percentiles.\\
As suggested by \citet{vogl2019individualized} for their measurements, contributions of higher frequencies in $Z\rm{_{ec}}$ were suppressed by reducing $f\rm{_{lim}}$ to 10\,kHz and the predictions of the transfer impedances for this database utilizing the method of the present study were repeated with an appropriately adjusted $f\rm{_{cut}}$ of 17.4\,kHz. However, as a result, the accuracy decreased, with a mean $L\rm{_{rmse}}$ of 4.77\,dB and a mean $\vartheta\rm{_{rmse}}$ of 23.65\,$^\circ$.

\newpage
\section{\label{sec:discussion} Discussion}
Utilizing the method of the present study with optimal, or re-calibrated, parameters on data obtained from three-dimensional FEM simulations for human ear canals and an ear canal simulator, the accuracy of the predicted ear canal area functions improves considerably in comparison to the method from \citet{rasetshwane2011inverse}. Nevertheless, a number of causes may impede the precision of the predicted area function.\\
Firstly, ear canals essentially resemble parabolically curved truncated cones with frequent tapering at the first bend. The cross-sections of real ear canals are elliptical rather than circular. However, the good agreement of the transfer impedance from the three-dimensional FEM simulation and that from the one-dimensional EA model fed with the vmtk area function of the ear canals imply that the neglected curvature of the centerline and the actual elliptical shape of the ear canal cross-sections probably contribute less to the error including frequencies between 1\,kHz and 10\,kHz (cf. $L\rm{_{rmse}}$ and $\theta\rm{_{rmse}}$ for 'ref. geometry' in Fig.~\ref{fig:box_error_sim}). A tendency to underestimate the area function still remains, especially at the lateral end of the ear canal and in regions of strongly varying area functions.\\
Secondly, the remaining underestimation of the area function might be due to incorrect assumptions about the sound field, which limit the range of validity of the theoretical calculations. For parabolic horn geometries with a perfect straight centerline and perfect circular cross-sections, the area functions show a similar underestimation to that seen for ear canal geometries, as could also be seen in \citet{rasetshwane2012reflectance} for the parabolic horn. An examination of the result for the ear canal simulator (Fig.~\ref{fig:neely3}, violet line) indicates that, with a more accurate estimate for $A(x=0)$, i.e., after adjusting $Z\rm{_0}(x=0)$, $A(x)$ would also be less underestimated in regions of strongly varying area functions.\\
Two different methods for adjusting the characteristic impedance $Z\rm{_0}(x=0)$, namely the one given by \citet{rasetshwane2011inverse} (conforming with \cite{rasetshwaneSourceCode2012old}, 'surge 1'), and the one
given in \citet{rasetshwaneSourceCode2012} ('surge 2'), were investigated. At the lateral end, the underestimation of $A(x=0)$ for ear canals appeared to vanish using the second method. On the other hand, the fitting resulting from the first method is preferable and even more suitable than the calculation of $Z\rm{_{0}}$ according to Eq.~(\ref{eq:Z0}) with the known geometric reference area. Dispersion and non-planar wave propagation, as well as strong changes in the area function, likely contribute to the error. As mentioned in Section~\ref{sec:intro}, an approach to determine a more accurate TDR for real ear canals can be found in \citet{norgaard2019calculation} and \citet{keefe2020sound}.\\
Thirdly, the inverse solution of Webster's horn equation is performed without additional boundary conditions at the eardrum i.e., by assuming a rigid termination. Also, rigid ear canal walls and lossless sound propagation are implied. This discrepancy between the simulated or measured input data and the model should also lead to an error that is not precisely quantified here. This phenomenon may be most apparent within the mid-frequency range around the minimum of the eardrum impedance level. Above 3\,kHz, the eardrum can be approximated as a rigid boundary \cite{hudde1999methods}.
In \citet{wulbusch2023using}, the eardrum impedance, coupled to a residual volume according to \citet{hudde1998measuringII} and \citet{hudde1999methods}, is assumed as a boundary condition at a length l, which is determined in the fitting process. The present study's methodology does not provide a fitting of this manner.\\
Lastly, the continuation of the input impedance spectrum, accompanied by the continuation of the reflectance spectrum, was found to be important in order to achieve an adequate spatial sampling through a high value of $f\rm{_{sup}}$ whereby the error can still be reduced to a relevant extent with a sampling $\Delta x$ of 0.1\,mm, according to \cite{xia2024effect}. Alternative approaches to model ear canal acoustics, i.e. \cite{sankowsky2011prediction} \cite{sankowsky2015individual} \cite{keefe2025shape}, may also benefit from a finer spatial resolution. But instead of extrapolating $Z\rm{_{ec}}$ to high frequencies (up to $f\rm{_{sup}}$/2) with a constant value, as proposed in the present work, there may be more suitable methods. A continuation of the input impedance spectrum predicted from lower frequencies ($<$$f\rm{_s}$/2) utilizing potential periodic properties of (non-)harmonic frequency components given by (non-)integer frequency ratios of resonances and antiresonances might be conceivable, as this would prevent the need for windowing.\\ 
The optimal ratio of $f\rm{_{cut}}$ to $f\rm{_{lim}}$ was examined using three-dimensional simulations for one sample of 21 ear canal geometries for three different lengths. Higher values of $f\rm{_{lim}}$ are recommended in order to increase the bandwidth of a favorable $f\rm{_{cut}}$ where the error rate is low (Fig.~\ref{fig:surf2}): For higher $f\rm{_{lim}}$, the method is more robust against a possibly non-optimal $f\rm{_{cut}}$ selection for other samples. In addition, it has been shown that the linear relationship is stable across subsamples and conditions, with limited variability in model parameters and moderate prediction error.
Furthermore, the additional error is not substantial for measurements of four different independent samples.\vspace{+0.5cm}\\
In the following, the suitability of the different lengths for the termination of the inverse solution for an optimal estimation of the transfer impedance is addressed.\\
The median distance between the umbo point (projected on the centerline) at $l\rm{_{umbo}}$ and the geometric ear canal end at $l\rm{_{geoend}}$ is 3.5\,mm, which is close to Hudde's assumption of a cone-like residual volume with a cone length of 4\,mm \cite{hudde1998measuringII}. The umbo point was used by \citet{rasetshwane2011inverse} as reflection point and is approximately given by the distance corresponding to the time at which the TDR reaches its maximum. $l\rm{_{TDRmax}}$ only slightly overestimates the reference umbo length in geometric ear canal models (mostly less than 2\,mm). The geometric end of the entire ear canal at $l\rm{_{geoend}}$ was associated with $l\rm{_{TDR50}}$ by \citet{rasetshwaneSourceCode2012}, which turned out to overestimate the geometric ear canal length to the same minor extent.\\
Regarding the termination of the inverse solution at $l\rm{_{lme}}$ for the most accurate prediction of $Z\rm{_{trans}}$, $l\rm{_{lme}}$ was found to coincide quite well with $l\rm{_{umbo}}$, with deviations less than 1\,mm in most cases. Since $l\rm{_{lme}}$ is unknown in the use case, the present study investigated a length criterion extractable from the available data. In addition to the termination criteria of mentioned lengths $l\rm{_{TDRmax}}$ and $l\rm{_{TDR50}}$, the global minimum of $\epsilon$ derived from the inverse solution at $l_{\epsilon}$, which reliably occurred between the projected umbo point and the innermost corner was proposed.
With length corrections that moved termination criteria closer to the length of the least mean error, small improvements for the level and considerable improvements for the predicted phase of $Z\rm{_{trans}}$ could be achieved in reference to simulated data for ear canal geometries. A termination of the inverse solution at $x$\,=\,$l_{\epsilon }\,-\,1.8\,mm$ or $x$\,=\,$l_{TDRmax}\,-\,0.9\,mm$ turned out to yield the most accurate $Z_{trans}$ for the three-dimensional FEM simulation.\\
However, regarding the validation results on measurements, the advantage of the length correction disappeared, presumably because correction lengths of 0.9\,mm and 1.8\,mm are smaller than the positioning accuracy of the probe tube at the eardrum, addressing the problem of an inaccurate reference.\vspace{+0.5cm}\\ 
Further aspects need to be considered with regard to the measurements. Especially in the database of \citet{vogl2019individualized}, it is frequently observed that the levels of $Z\rm{_{trans}}$ given in the database are lower than levels of corresponding $Z\rm{_{ec}}$, a phenomenon that did not occur in the simulations. Therefore, a bias to positive level errors around 4-6\,kHz was expected but did not occur for the data from \citet{vogl2019individualized}. Conversely, a tendency to negative errors around 2.5\,kHz (slightly above the first, strongly flattened anti-resonance in $Z\rm{_{ec}}$) can presumably be explained by particularly unfavorable geometries, as the input impedance is not determined at the residual ear canal, but in the vent of the ear mold. Thus part of the ear mold had to be accommodated by the area function, with a cross-sectional jump at the entrance to the residual ear canal, violating the assumptions of Webster's horn equation.\\
Contrary to expectations, reducing $f\rm{_{lim}}$ to 10\,kHz and $f\rm{_{cut}}$ to 17.4\,kHz for this database in order to weaken potentially non-valid contributions of higher frequencies in the reflectance decreased the prediction accuracy for $Z\rm{_{trans}}$, confirming the benefit of higher $f\rm{_{lim}}$.\\
Generally, transfer impedance errors resulting from application of the (optimized) method of the present study did not exceed tolerance limits as defined in related publications between 1 and 10\,kHz, or can be attributed to measurement inaccuracies as addressed in Section~\ref{sec:meas}. Below 1\,kHz, errors were larger, but also less relevant, as the desired transfer impedance equals the known input impedance at these frequencies. Reported errors in related publications of the databases from \citet{blau2010prediction}, \citet{sankowsky2011prediction} and \citet{sankowsky2015individual} are only slightly smaller. For the database from \citet{vogl2019individualized}, the method presented there achieves smaller level errors, in particular in the medium frequency range. For a thorough investigation, a systematic comparison of all methods to predict $Z\rm{_{trans}}$ is necessary.\\

\section{\label{sec:conclusion} Conclusion}
Predictions of ear canal area functions from measured input impedances can be substantially improved by increasing the upsampling frequency to 3.5\,MHz, corresponding to 0.1\,mm spatial resolution at the CFL limit, by increasing the frequency range of the input impedance of the residual ear canal and by adjusting the cut-off frequency of the Blackman window depending on the highest frequency of the input impedance using the linear regression model as presented. Slight and systematic underestimations remain for parabolic shapes, probably due to the limitations of Webster's horn equation. For a sufficiently high spatial resolution, the global minimum of the gradient of the logarithmic inverse solution $\epsilon$ and the maximum of the time domain reflectance are both good predictors of the ear canal length.
For the purpose of individual equalization of the eardrum sound pressure, the method can be supplemented by a one-dimensional electro-acoustic model capable of predicting the transfer impedance of the ear canal. An equalization based on the method of present study presumably outperforms non-individualized equalization algorithms, but needs to be compared to other methods.\\
The advantage of the method, despite slight underestimates of the area functions, is its reproducibility and robustness with consistently plausible results for a wide range of insertion depths. Finally, the computational load is very low compared to non-linear fitting algorithms. The inverse solution could be employed to initialize and constrain parameters of other fitting algorithms making them more robust.

\begin{acknowledgments}
	We would like to thank the anonymous reviewers for
	helpful comments on an earlier version of the manuscript.\\
	\textbf{Funding}\\
	This research was partially funded by the Deutsche Forschungsgemeinschaft (DFG, German Research
	Foundation)—Project ID 352015383—SFB 1330 C1.
\end{acknowledgments}

\section*{\label{sec:declaration}AUTHOR DECLARATIONS}
\vspace*{-0.25cm}
\subsection*{Conflict of Interest}
The authors have no conflicts to disclose.

\section*{\label{sec:data}DATA AVAILABILITY}
The data that support the findings of this study are available from the corresponding author upon reasonable request.

\bibliography{bib_inverse_solution}

@article{strutt1871some,
	title={Some general theorems relating to vibrations},
	author={Strutt, John William},
	journal={Proceedings of the London Mathematical Society},
	volume={1},
	number={1},
	pages={357--368},
	year={1871},
	publisher={Oxford University Press}
}

@article{webster1919acoustical,
	title={Acoustical impedance and the theory of horns and of the phonograph},
	author={Webster, Arthur Gordon},
	journal={Proceedings of the National Academy of Sciences},
	volume={5},
	number={7},
	pages={275--282},
	year={1919},
	publisher={National Acad Sciences}
}

@article{courant1928partiellen,
	title={{\"U}ber die partiellen {D}ifferenzengleichungen der mathematischen {P}hysik},
	author={Courant, Richard and Friedrichs, Kurt and Lewy, Hans},
	journal={{M}athematische {A}nnalen},
	volume={100},
	number={1},
	pages={32--74},
	year={1928},
	publisher={Springer}
}

@article{goldsmith1924performance,
	title={The performance and theory of loud speaker horns},
	author={Goldsmith, Alfred N and Minton, John P},
	journal={Proceedings of the Institute of Radio Engineers},
	volume={12},
	number={4},
	pages={423--478},
	year={1924},
	publisher={IEEE}
}

@article{benade1968propagation,
  title={On the propagation of sound waves in a cylindrical conduit},
  author={Benade, Arthur H},
  journal={The Journal of the Acoustical Society of America},
  volume={44},
  number={2},
  pages={616--623},
  year={1968},
  publisher={AIP Publishing}
}

@article{haakansson1986mechanical,
	title={The mechanical point impedance of the human head, with and without skin penetration},
	author={H{\aa}kansson, Bo and Carlsson, Peder and Tjellstr{\"o}m, Anders},
	journal={The Journal of the Acoustical Society of America},
	volume={80},
	number={4},
	pages={1065--1075},
	year={1986},
	publisher={Acoustical Society of America}
}

@article{stinson1989specification,
	title={Specification of the geometry of the human ear canal for the prediction of sound-pressure level distribution},
	author={Stinson, Michael R and Lawton, BW},
	journal={The Journal of the Acoustical Society of America},
	volume={85},
	number={6},
	pages={2492--2503},
	year={1989},
	publisher={Acoustical Society of America}
}

@article{keefe1992method,
	title={Method to measure acoustic impedance and reflection coefficient},
	author={Keefe, Douglas H and Ling, Robert and Bulen, Jay C},
	journal={The Journal of the Acoustical Society of America},
	volume={91},
	number={1},
	pages={470--485},
	year={1992},
	publisher={Acoustical Society of America}
}

@article{voss1994measurement,
  title={Measurement of acoustic impedance and reflectance in the human ear canal},
  author={Voss, Susan E and Allen, Jont B},
  journal={The Journal of the Acoustical Society of America},
  volume={95},
  number={1},
  pages={372--384},
  year={1994},
  publisher={Acoustical Society of America}
}

@article{hudde1998measuringII,
	title={Measuring and modeling basic properties of the human middle ear and ear canal. Part II: Ear canal, middle ear cavities, eardrum, and ossicles},
	author={Hudde, H and Engel, A},
	journal={Acta Acustica united with Acustica},
	volume={84},
	number={5},
	pages={894--913},
	year={1998},
	publisher={European Acoustics Association}
}

@article{hudde1999methods,
	title={Methods for estimating the sound pressure at the eardrum},
	author={Hudde, H and Engel, A and Lodwig, A},
	journal={The Journal of the Acoustical Society of America},
	volume={106},
	number={4},
	pages={1977--1992},
	year={1999},
	publisher={Acoustical Society of America}
}

@phdthesis{antiga2002patient,
	title={Patient-specific modeling of geometry and blood flow in large arteries},
	author={Antiga, Luca},
	school={Politecnico di Milano},
	year={2002}
}

@book{kuttruff2007acoustics,
	title={Acoustics: an introduction},
	author={Kuttruff, Heinrich},
	year={2007},
	publisher={CRC Press}
}

@article{blau2010prediction,
	title={Prediction of the sound pressure at the ear drum in occluded human cadaver ears},
	author={Blau, Matthias and Sankowsky, Tobias and Roeske, Philipp and Mojallal, Hamidreza and Teschner, Magnus and Thiele, Cornelia},
	journal={Acta Acustica united with Acustica},
	volume={96},
	number={3},
	pages={554--566},
	year={2010},
	publisher={European Acoustics Association}
}

@article{sankowsky2011prediction,
	title={Prediction of the sound pressure at the ear drum in occluded human ears},
	author={Sankowsky-Rothe, Tobias and Blau, Matthias and Rasumow, Eugen and Mojallal, Hamidreza and Teschner, Magnus and Thiele, Cornelia},
	journal={Acta acustica united with Acustica},
	volume={97},
	number={4},
	pages={656--668},
	year={2011},
	publisher={European Acoustics Association}
}

@article{rasetshwane2011inverse,
	title={Inverse solution of ear-canal area function from reflectance},
	author={Rasetshwane, Daniel M and Neely, Stephen T},
	journal={The Journal of the Acoustical Society of America},
	volume={130},
	number={6},
	pages={3873--3881},
	year={2011},
	publisher={AIP Publishing}
}

@article{rasetshwane2012reflectance,
	title={Reflectance of acoustic horns and solution of the inverse problem},
	author={Rasetshwane, Daniel M and Neely, Stephen T and Allen, Jont B and Shera, Christopher A},
	journal={The Journal of the Acoustical Society of America},
	volume={131},
	number={3},
	pages={1863--1873},
	year={2012},
	publisher={AIP Publishing}
}

@inproceedings{sankowsky2012prediction,
	title={Prediction of the sound pressure at the ear drum for open fittings},
	author={Sankowsky-Rothe, Tobias and Blau, Matthias and Mojallal, Hamidreza and Teschner, Magnus and Thiele, Cornelia},
	booktitle={Acoustics 2012},
	year={2012}
}

@article{sankowsky2015individual,
	title={Individual equalization of hearing aids with integrated ear canal microphones},
	author={Sankowsky-Rothe, T and Blau, M and K{\"o}hler, S and Stirnemann, A},
	journal={Acta Acustica united with Acustica},
	volume={101},
	number={3},
	pages={552--566},
	year={2015},
	publisher={European Acoustics Association}
}

@article{lewis2015non,
	title={Non-invasive estimation of middle-ear input impedance and efficiency},
	author={Lewis, James D and Neely, Stephen T},
	journal={The Journal of the Acoustical Society of America},
	volume={138},
	number={2},
	pages={977--993},
	year={2015},
	publisher={AIP Publishing}
}

@article{berggren2018acoustic,
	title={Acoustic boundary layers as boundary conditions},
	author={Berggren, Martin and Bernland, Anders and Noreland, Daniel},
	journal={Journal of Computational Physics},
	volume={371},
	pages={633--650},
	year={2018},
	publisher={Elsevier}
}

@article{bach2018theory,
	title={Theory of pressure acoustics with viscous boundary layers and streaming in curved elastic cavities},
	author={Bach, Jacob S and Bruus, Henrik},
	journal={The Journal of the Acoustical Society of America},
	volume={144},
	number={2},
	pages={766--784},
	year={2018},
	publisher={AIP Publishing}
}

@article{vogl2019individualized,
	title={Individualized prediction of the sound pressure at the eardrum for an earpiece with integrated receivers and microphones},
	author={Vogl, Steffen and Blau, Matthias},
	journal={The Journal of the Acoustical Society of America},
	volume={145},
	number={2},
	pages={917--930},
	year={2019},
	publisher={AIP Publishing}
}

@article{norgaard2019calculation,
	title={On the calculation of reflectance in non-uniform ear canals},
	author={N{\o}rgaard, Kren Rahbek and Charaziak, Karolina K and Shera, Christopher A},
	journal={The Journal of the Acoustical Society of America},
	volume={146},
	number={2},
	pages={1464--1474},
	year={2019},
	publisher={AIP Publishing}
}

@article{keefe2020sound,
	title={Sound field estimation near the tympanic membrane using area-distance measurements in the ear canal},
	author={Keefe, Douglas H},
	journal={The Journal of the Acoustical Society of America},
	volume={148},
	number={3},
	pages={1193--1214},
	year={2020},
	publisher={Acoustical Society of America}
}

@inproceedings{roden2020iha,
	title={The {IHA} database of human geometries including torso, head and complete outer ears for acoustic research},
	author={Roden, Reinhild and Blau, Matthias},
	booktitle={INTER-NOISE and NOISE-CON Congress and Conference Proceedings},
	volume={261},
	number={2},
	pages={4226--4237},
	year={2020},
	organization={Institute of Noise Control Engineering}
}

@misc{roden2021iha,
	author={Roden, Reinhild and Blau, Matthias},
	title     = {The {IHA} database of human geometries
	including torso, head and complete outer ears for acoustic research},
	version   = {1.0},
	publisher = {Zenodo},
	year      = {2021},
	doi       = {10.5281/zenodo.5528765},
	url       = {https://doi.org/10.5281/zenodo.5528765}
}

@article{balouch2023measurements,
	title={Measurements of ear-canal geometry from high-resolution CT scans of human adult ears},
	author={Balouch, Auden P and Bekhazi, Karen and Durkee, Hannah E and Farrar, Rebecca M and Sok, Mealaktey and Keefe, Douglas H and Remenschneider, Aaron K and Horton, Nicholas J and Voss, Susan E},
	journal={Hearing Research},
	volume={434},
	pages={108782},
	year={2023},
	publisher={Elsevier}
}

@article{wulbusch2023using,
	title={Using a one-dimensional finite-element approximation of Webster's horn equation to estimate individual ear canal acoustic transfer from input impedances},
	author={Wulbusch, Nick and Roden, Reinhild and Chernov, Alexey and Blau, Matthias},
	journal={The Journal of the Acoustical Society of America},
	volume={153},
	number={5},
	pages={2826--2826},
	year={2023},
	publisher={AIP Publishing}
}

@article{xia2024effect,
	title={Effect of curvature on sound propagation in the ear canal},
	author={Xia, Yuanxin and Guo, Zhihan and Tiana-Roig, Elisabet and Henriquez, Vicente Cutanda and Lucklum, Frieder},
	journal={The Journal of the Acoustical Society of America},
	volume={155},
	number={1},
	pages={695--706},
	year={2024},
	publisher={AIP Publishing}
}

@article{keefe2025shape,
	title={Shape and sound analyses of the human ear-canal geometry},
	author={Keefe, Douglas H and Porter, Heather L and Fitzpatrick, Denis F},
	journal={The Journal of the Acoustical Society of America},
	volume={157},
	number={5},
	pages={3638--3654},
	year={2025},
	publisher={AIP Publishing}
}

@misc{rasetshwaneSourceCode2012,
	author = {Daniel Rasetshwane},
	year = {2012},
	note = {Reflectance measurements and solution to the inverse problem (version 2) at\\
	\url{http://audres.org/cel/refl/measures.html},
	(Last viewed August 25, 2022)}}

@misc{rasetshwaneSourceCode2012old,
author = {Daniel Rasetshwane},
year = {2012},
note = {Reflectance measurements and solution to the inverse problem (version 1) at\\
\url{http://audres.org/cel/refl/measures.html},
(Last viewed January 26, 2012)}}

@book{hastie2009elements,
	title={The elements of statistical learning},
	author={Hastie, Trevor and Tibshirani, Robert and Friedman, Jerome and others},
	year={2009},
	publisher={Springer series in statistics New-York},
	pages ={ 241--260}
}

\end{document}